\documentclass[useAMS,usenatbib]{mn2e}
\usepackage{pifont}
\usepackage{graphicx}
\usepackage{multirow}
\usepackage{textcomp}
\usepackage{amssymb}
\usepackage{color}
\usepackage{abbrevs}
\usepackage{xspace}
\usepackage{float}

\begingroup
  \catcode`\_=\active
  \gdef_#1{\ensuremath{\sb{\mathrm{#1}}}}
\endgroup
\mathcode`\_=\string"8000
\catcode`\_=12

\newcommand{\Msolar}{M$_{\odot}$\xspace}
\newcommand{\Zsolar}{Z$_{\odot\xspace}$\xspace}

\newcommand{\atcc}{atoms/cm$^{3}$\xspace}

\newcommand{\simname}{\textsc}
\newcommand{\Stromgren}{Str\"{o}mgren}

\newcommand{\tick}{\ding{51}}

\newcommand{\rev}{\textcolor{black}}

\newabbrev\ISM{Interstellar Medium (ISM)}[ISM]
\newabbrev\CSM{Circumstellar Medium (CSM)}[CSM]
\newabbrev\WNM{Warm Neutral Medium (WNM)}[WNM]
\newabbrev\WIM{Warm Ionised Medium (WIM)}[WIM]
\newabbrev\CNM{Cold Neutral Medium (CNM)}[CNM]
\newabbrev\IMF{Initial Mass Function (IMF)}[IMF]
\newabbrev\AMR{Adaptive Mesh Refinement (AMR)}[AMR]
\newabbrev\HGB{Horizontal Giant Branch (HGB)}[HGB]

\makeatletter
\renewcommand\maybe@space@{%
  \maybe@ictrue 
  \expandafter   \@tfor
    \expandafter \reserved@a
    \expandafter :%
    \expandafter =%
                 \nospacelist
                 \do \t@st@ic
  \ifmaybe@ic 
    \space
  \fi
}
\makeatother

\begin{document}
\title{A Detailed Study of Feedback from a Massive Star}
\author[S. Geen]
      {Sam Geen$^{1, 2}$, Joakim Rosdahl$^{2,3}$, Jeremy Blaizot$^{2}$, Julien Devriendt$^{2,4}$, Adrianne Slyz$^{4}$\\
{$^{1}$Laboratoire AIM, Paris-Saclay, CEA/IRFU/SAp - CNRS - Universit\'e Paris Diderot, 91191, Gif-sur-Yvette Cedex, France}\\
{$^{2}$CRAL, Universit\'e de Lyon I, CNRS UMR 5574, ENS-Lyon,  9 avenue Charles Andr\'e, 69561 Saint-Genis-Laval, France}\\
{$^{3}$Leiden Observatory, Leiden University, P.O. Box 9513, 2300 RA, Leiden, The Netherlands}\\
{$^{4}$University of Oxford, Astrophysics, Keble Road, Oxford OX1 3RH, UK}\\}
\date{\today}
\maketitle

\begin{abstract}
We present numerical simulations of a 15 \Msolar star in a suite of idealised environments in order to quantify the amount of energy transmitted to the interstellar medium (ISM). We include models of stellar winds, UV photoionisation and the subsequent supernova based on theoretical models and observations of stellar evolution. The system is simulated in 3D using RAMSES-RT, an Adaptive Mesh Refinement Radiation Hydrodynamics code. We find that stellar winds have a negligible impact on the system owing to their relatively low luminosity compared to the other processes. The main impact of photoionisation is to reduce the density of the medium into which the supernova explodes, reducing the rate of radiative cooling of the subsequent supernova. Finally, we present a grid of models quantifying the energy and momentum of the system that can be used to motivate simulations of feedback in the ISM unable to fully resolve the processes discussed in this work.
\end{abstract}

\begin{keywords}
(stars:) massive, (ISM:) H ii regions, supernova remnants, (nebulae, stars:) winds, outflows, (supernovae:) general, (methods:) numerical 
\end{keywords}


\section{Introduction}
\label{introduction}



%
%
%
%
%
%
%

The $\Lambda$CDM (Cosmological Constant $\Lambda$ + Cold Dark Matter) model of cosmological galaxy formation has achieved great successes in explaining the large-scale structure of the universe. However, problems remain with this model that must be addressed before we can match completely our theoretical models with observations. One key issue is that the stellar masses of galaxies observed in the universe are lower than what would be expected if each galaxy were embedded in a dark matter halo with a constant luminosity-halo mass proportion. Similarly, large-scale mass outflows from galaxies have been observed \citep[see review by][]{Veilleux2005}. In both cases, stellar feedback, particularly from supernovae, but also from stellar photoionisation and winds, has been employed with varying success to explain this discrepancy\footnote{Stellar feedback refers to the ability of stellar evolution processes to regulate the subsequent star formation rate}. Simple analytical and semi-analytical models (analytical 
models run in a framework of pure dark matter cosmological N-body simulations) have found that the energy from stellar sources is sufficient to launch galactic winds and suppress star formation in lower-mass halos \citep{Benson:2003p1159}.

Despite the success of recent models in explaining many of the observed properties of galaxies, hydrodynamical numerical simulations of galaxy formation have encountered difficulties in reproducing these results in more self-consistent settings. \cite{ScannapiecoC2011} find that there is still significant disagreement between analytical models, SAMs, and hydrodynamic simulations, both smoothed-particle hydrodynamics (SPH) and adaptive-mesh refinement (AMR) simulations. A large part of this problem is the limitation imposed by numerical resolution. If adequate numerical resolution is not achieved, the gas will cool radiatively before it has a chance to add momentum to the \ISM. \cite{GerritsenJ.P.E.1997} address this problem by enforcing thermal equilibrium for gas particles with a cooling time of less than 10\% of the current timestep. \cite{Hopkins2013} propose instead to deposit the supernova blast onto the grid as momentum if the grid resolution is below the cooling length as calculated by \cite{Cox1972}, using values for momentum calibrated elsewhere. \rev{A similar method has been implemented by \cite{Kimm2014}. \cite{Iffrig2014} find that the mometum added to the \ISM can be well approximated by taking the momentum of a Sedov blastwave \citep{Sedov1946} at the cooling time given by \cite{Cox1972}.}

A major problem faced by numerical simulations of galaxy formation is to understand how this energy created by stars accounts for the observed properties of gas in the \ISM and galactic outflows. This problem is complicated by the fact that the \ISM is a multiphase medium that much be simulated with resolutions on the scale of parsecs or below if we wish to capture it without resorting to sub-grid modelling (a sub-grid model is an expression implemented in the code to account for processes that cannot be spatially resolved by the simulation). The densest phase is the \CNM, made up of clouds and filaments at around 100K. These are embedded in a warm, diffuse phase at around 10$^4$ K called the \WNM if neutral or \WIM if ionised by UV radiation. 
A hot ionised medium at above 10$^6$ K exists in bubbles formed by supernova explosions.
As proposed in the results of analytical models by \cite{McKee:1977p1412}, the presence of these phases is thought to be the result of multiple supernova explosions.
\cite{Springel:2003p1820,Murante:2010p1836} attempt to circumvent the limitations of resolution in their simulations with a ``sub-grid'' model for the interaction between the cold and hot gas phases that are traced separately inside each fluid element. \cite{Springel:2003p1820} also invoke winds phenomenologically to allow the escape of hot gas from the galaxy without resolving the evolution of supernova remnants in the \ISM. Meanwhile, \cite{NavarroJ.F.1993,Mihos1994} model stellar feedback using kinetic winds. \cite{Dubois2008} account for a lack of resolution by imposing a Sedov profile onto the gas when a supernova occurs.

The lifetime of an OB star is of order 10 Myr. The precise age depends on a number of factors such as mass, chemical composition and rotation velocity, as well as multiplicity, i.e. interactions with a companion star or stellar remnant. For the purposes of this paper we ignore binary supernovae such as Type Ia, since the lifetimes of their progenitors are much longer and as such do not induce such immediate feedback into the \ISM as single-star Type II supernovae, though ultimately their energy contribution may be important. \cite{Heger2003} state that stars must be over 8-10 \Msolar to explode as supernovae. Further, they argue that above around 25 \Msolar the type of supernova depends on the mass and metallicity, with very massive low-metallicity stars undergoing direct collapse to a black hole, many with a weak or no supernova. As stars of lower masses are more common according to the standard IMFs proposed by \cite{Salpeter:1955p2059,Chabrier2003,Kroupa:2003p2022}, the energy budget from supernovae of stars above 25 \Msolar must be less than that of stars between 9 and 25 \Msolar. Supernovae at the lower end 
of this range release approximately 10$^{51}$ ergs as kinetic energy into the \CSM \citep{Chevalier:1977p1817}, though \cite{Nomoto2003} suggest that more massive stars exploding as hypernovae can release up to around 50 times this value, noting that, depending on its composition and evolution, a star above 25 \Msolar can also produce a faint (below 10$^{51}$ ergs) supernova.

Estimates based on 1D simulations in a uniform medium suggest that only 3 to 10\% of the 10$^{51}$ ergs of kinetic energy produced by a supernova is transferred to the \ISM depending on the physics modelled and the density of the external medium, and the remaining energy is lost to thermal radiation \citep{Chevalier1974,SpitzerLyman1978}. Initially, once the shock has broken out of the star, it evolves adiabatically according to the Sedov-Taylor solution \citep{Sedov1946}. Once the supernova remnant's thermal energy falls below its kinetic energy, it enters a pressure-driven snowplough phase. At this point, the pressure force from the hot, diffuse gas inside the remnant drops so that it is comparable to the deceleration from the accretion of matter from the external medium by the cold, dense shell surrounding the remnant. As the thermal pressure inside the remnant drops further, it enters a momentum-conserving snowplough phase. Eventually, the remnant is disrupted and destroyed when it merges with the turbulent \ISM surrounding it.
\cite{Cioffi1988} produce analytic empirically$-$motivated models of the evolution of a supernova remnant and estimate this merging time, which can in certain circumstances happen before the momentum-conserving phase is reached. A parameterised study of a single supernova in various uniform media was performed by \cite{Thornton1998}, who find that while a remnant cools faster when the supernova explodes in a denser medium, the resulting kinetic energy in the dense shell is remarkably constant with the external medium's density and metallicity. Highly diffuse media can produce highly adiabatic shocks \citep{Tang:2005p1411}, whilst very dense media produce supernova remnants 
that become momentum-conserving almost immediately \citep{Tenorio-TagleG.1991}. More recently, \cite{Martizzi2014,Iffrig2014} have studied isolated supernovae in multiphase environments in an ettempt to address this issue. A further question is whether massive stars explode in dense clouds at all. \cite{Slyz2005} argue that the delay between star formation and the first supernova (given by the lifetime of massive stars as discussed above) enhances the multiphase \ISM and star formation rates by allowing massive stars to drift out of star-forming clouds into lower density regions before exploding. \cite{Ceverino2009,Kimm2014,Hennebelle2014} find that ``runaway stars'' can drastically reduce energy loss rates from supernovae, produce more realistic galaxy bulge masses and increase the escape fraction of UV photons, since the supernovae now explode outside dense star-forming environments. 

The impact of pre-supernova stellar feedback can also play a role in injecting energy into the \ISM and modifying the environment into which supernovae explode. Star formation occurs in the \CNM, where the gas is Jeans-unstable and collapses to form star-forming cores in molecular clouds. Stars feed back into this environment via three main processes - UV photoionisation, stellar winds and protostellar jets. Early work by \cite{Stromgren1939,KahnF.D.1954,Oort1955} argues that radiation feedback by UV photons emitted by OB stars plays an important role in regulating star formation in clouds. These photons heat the gas in clouds to around 10$^4$ K, preventing further star formation and drive thermal shocks that expel gas from the clouds. This is explored in simulations by \cite{Dale2005,Arthur2011,Walch2012,Walch2013,Dale2014}, and the observations of, e.g., \cite{Chu1994,Redman2003}. On a smaller scale, \cite{Bate2012} argue that radiation feedback plays an important role in 
regulating the formation of star-forming cores and hence the shape of the \IMF. 

\cite{Krumholz2009} propose that radiation pressure may play a role in driving feedback from OB stars. However, we do not consider radiation pressure in this work. \cite{Krumholz2012c,Sales2014} and Rosdahl et al. (2015, in prep) conclude that the impact of radiation pressure compared to that of UV photoheating is limited, though there may be regimes in which it becomes important.

\cite{Castor1975,Avedisova1972,Weaver1977} produce analytic expressions for the evolution of stellar wind-driven bubbles in the adiabatic regime. Unlike ionisation fronts, which produce shocks via thermal differences between the ionised and neutral gas, stellar winds produce shocks via the interaction of winds travelling on the order of the escape velocity of the star \citep{Kudritzki2000} and the circumstellar medium. The balance of available energy from either process depends on the properties of the star. Higher metallicity stars are more opaque, and thus have lower luminosities whilst driving stronger stellar winds, whilst low mass stars may not produce enough UV photons to ionise the surrounding medium. Recent work by \cite{Dale2014} has explored the relative impact of winds and photoionisation from young star clusters on molecular cloud evolution. Working on larger scales, \cite{Agertz2013} produces a sub-grid model for galaxy formation simulations that gives the energy produced by each feedback 
process from a population of stars. Jets from protostars could also help explain low star formation efficiencies in star-forming clouds. See the review by \cite{Krumholz2014} for more on this subject. These are mainly of importance in young clusters with active star formation, and as such will be implemented in future work studying feedback in these environments.

Considerable work has been carried out already on the evolution of supernovae inside circumstellar media previously modified by stellar winds and photoionisation. Indeed, diverse structures in supernova remnants have been attributed to the existing density structure of the \CSM, which is often created by the supernova progenitor prior to the explosion. \cite{Dwarkadas2007} uses numerical simulations to explain observed structures in supernova remnants by invoking a Wolf-Rayet wind prior to the supernova, while \cite{Walch2014} discuss the interaction between a pre-existing photoionised cloud and a supernova explosion. Pre-supernova stellar feedback can also alter the geometry of the supernova remnant. \cite{GarciaSegura1999} note that stellar rotation can induce a bipolar structure in the wind-blown \CSM. \cite{VanMarle2008} argue that stellar rotation causes the density profile of the \CSM to diverge from a \cite{Chevalier1982} power law.  \cite{Tenorio-TagleG.1990,RozyczkaM.1993} suggest that stellar motion with respect to the \ISM gas can produce barrel-shaped supernova remnants as pre-supernova winds carve out a tube-like structure in the \ISM, while \cite{Mackey2014} explore the interaction between the wind and ionisation front in the context of a star moving with respect to the \CSM. Supernova shocks are subject to turbulence driven by Rayleigh-Taylor, Vishniac and, in the case of non-spherical shocks, Kelvin-Helmholtz instabilties. \cite{GullS.F.1973} propose that these instabilities can modify the energetics of a supernova shock by converting kinetic energy on the shock into thermal energy via the turbulent energy cascade. Numerical simulations by \cite{Dwarkadas2007,Fraschetti2010a} report the growth of these instabilities. \rev{\cite{Ntormousi2011,Krause2012}} find that stellar wind shock fronts are also unstable, and determine that wind-blown bubbles will be prone to Vishniac instabilities, which grow due to radiative cooling and self-gravity \rev{(not included in our simulations)} \citep{Vishniac1983,Vishniac1994}. By contrast, \cite{Ricotti2014} argue that ionisation fronts are not typically turbulent.

The role of this paper is to update the work of \cite{Thornton1998} by taking into account the role of photoionisation and stellar winds from a single 15\Msolar star on the evolution of its subsequent supernova remnant in a set of uniform media of various densities and metallicities. In addition to this, we take advantage of advances in computing to run 3D, rather than spherically-symmetric 1D simulations as in, e.g. \cite{Chevalier1974,Cioffi1988,Thornton1998}. The advantage of using 3D simulations as opposed to 1D simulations is that we are able to quantify the impact of these instabilities on the energetics of the supernova. We resolve the gas to sub-parsec resolutions such that our results converge in test runs (see section \ref{methods:numsim}).

Our paper is organised as follows. Section \ref{methods} is concerned with the models used for stellar winds, photoionisation and supernova feedback, as well as the setup of the numerical simulations. In Section \ref{evolution} we present in detail one of our simulations in order to give a qualitative description of the structures formed by the star. We then look at the response of two sample environments to winds and photoionisation by studying each process in isolation. Section \ref{supernova} discusses how including each of the processes affects the energy and momentum transferrred to media at various densities and metallicites. Finally, we discuss our results and some possible limitations in light of simplifications made by the study.


\section{Methods}
\label{methods}

\subsection{A Model Star}
\label{methods:model}

%
%
%
%
%
%
%

\begin{figure}
\centerline{\includegraphics[width=1.0\hsize]{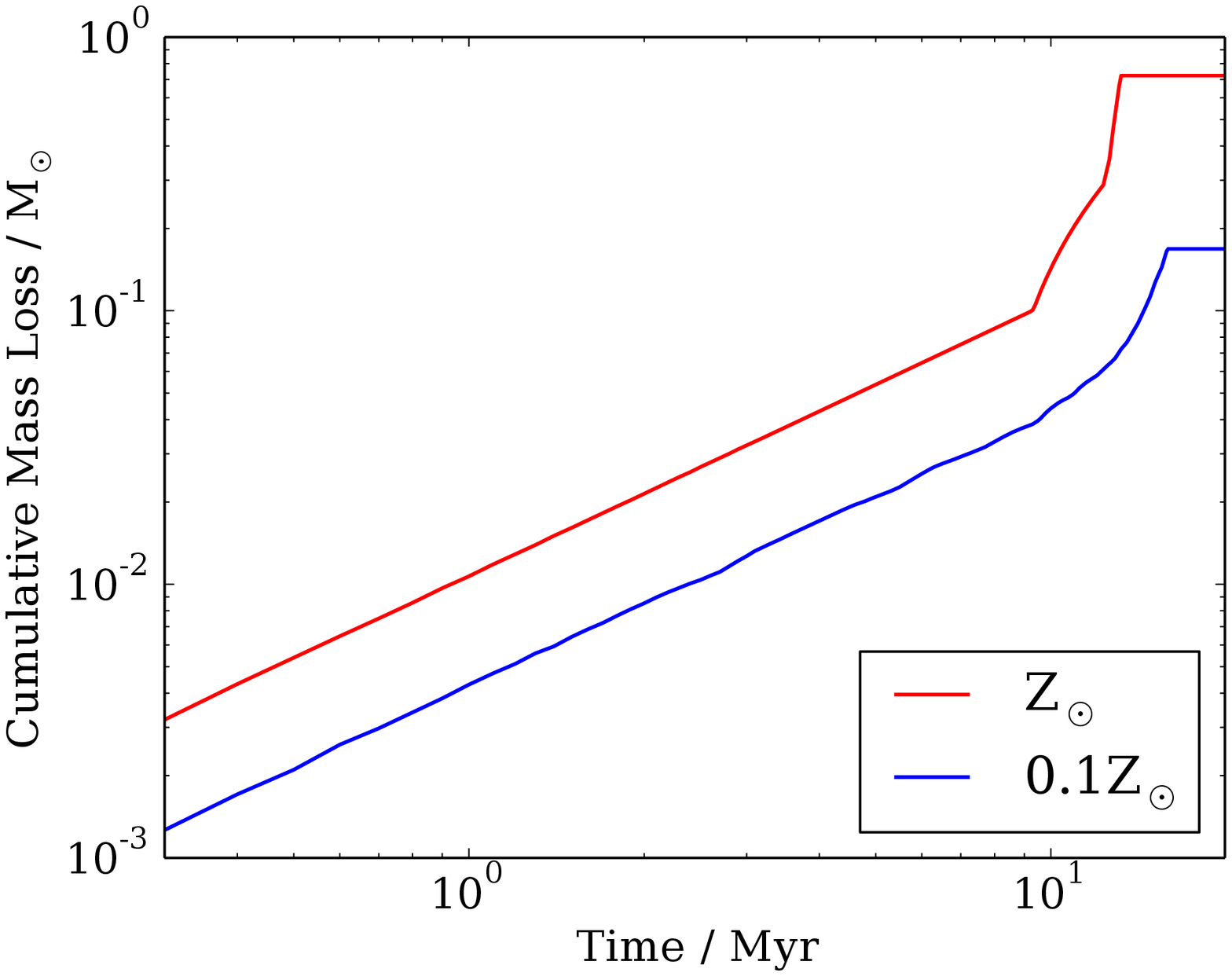}}
\centerline{\includegraphics[width=1.0\hsize]{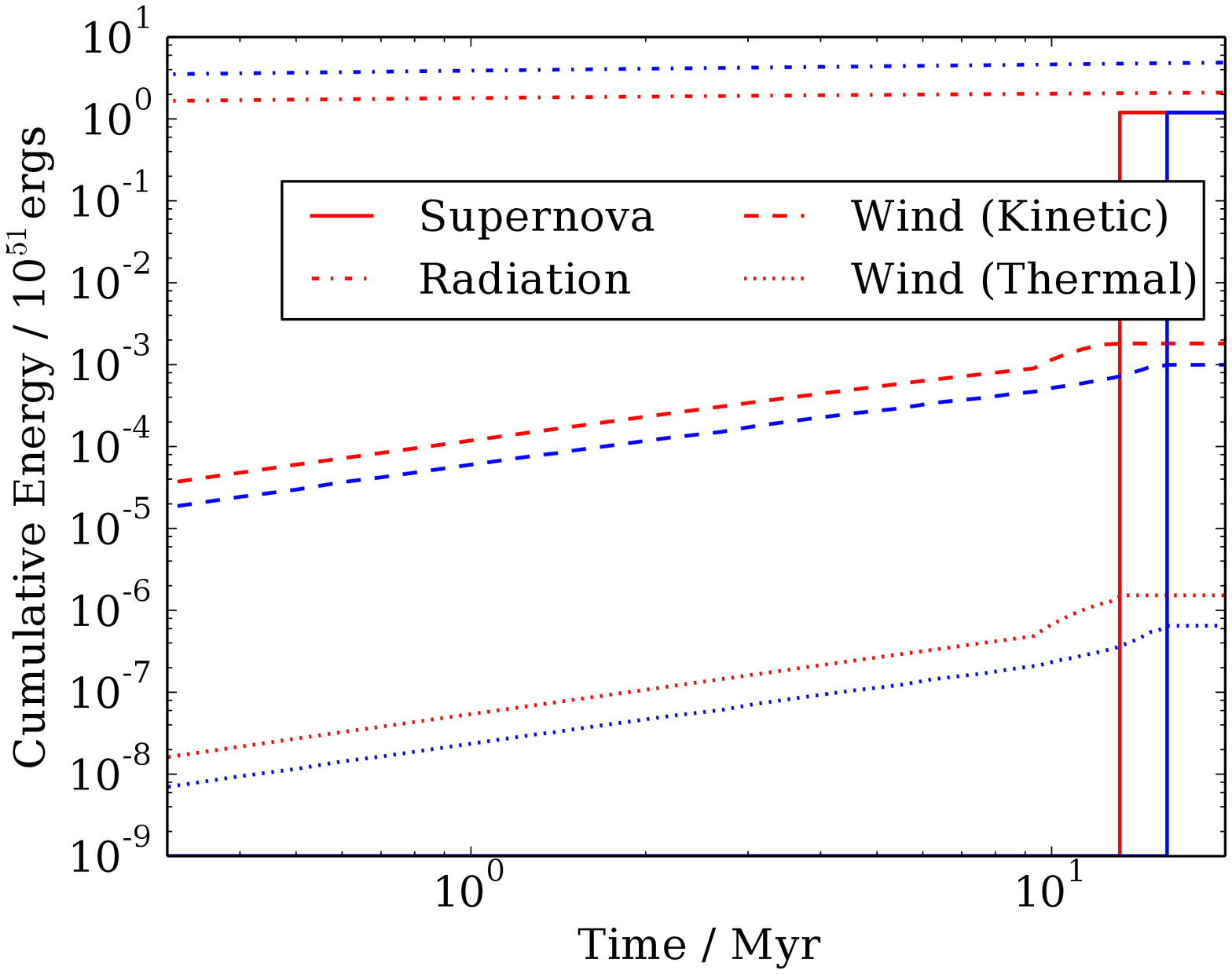}}
\centerline{\includegraphics[width=1.0\hsize]{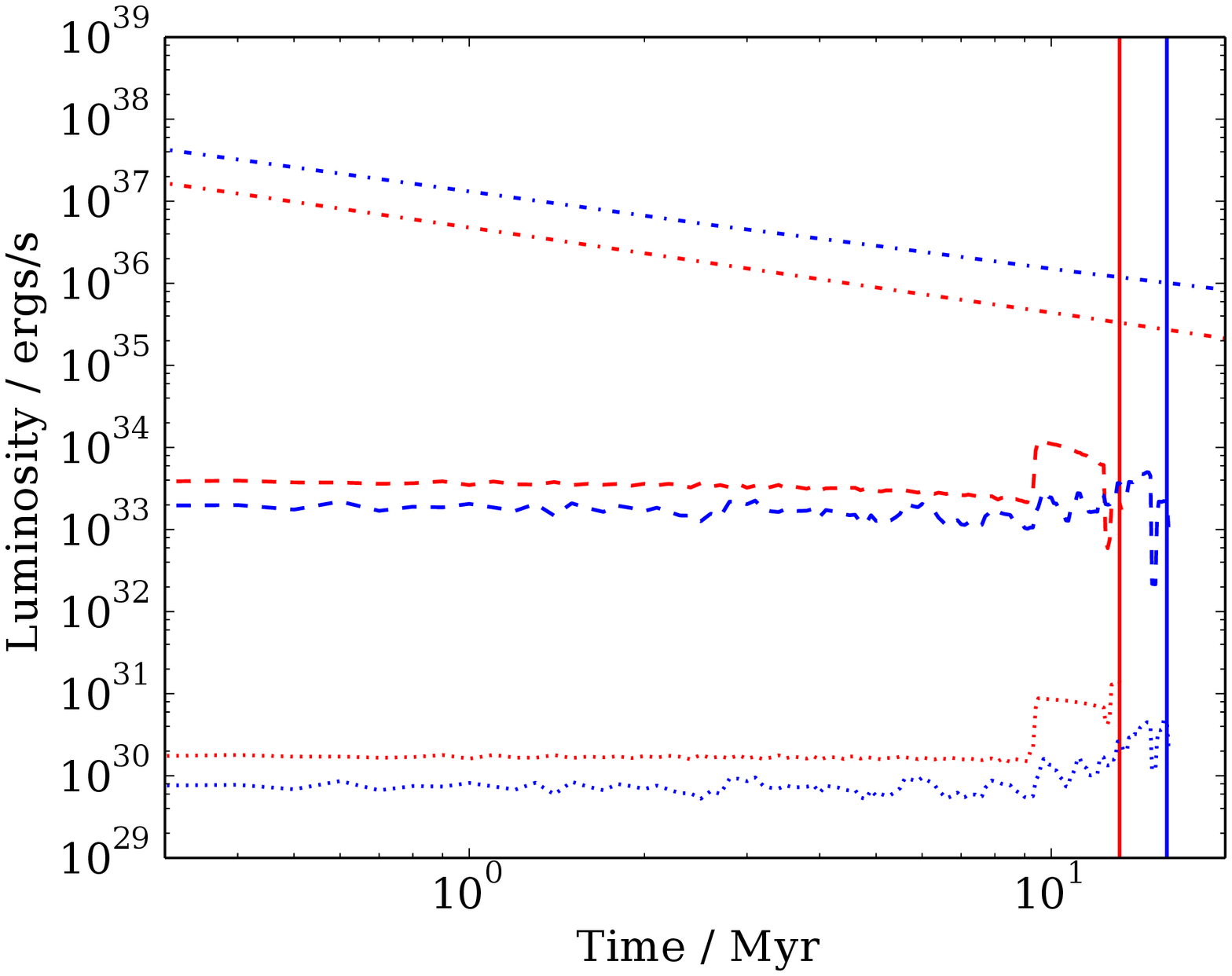}}
  \caption{Cumulative mass loss (top), cumulative energy output (middle) and energy output rate (bottom) against time from a 15 \Msolar star of metallicities \Zsolar in red and 0.1 \Zsolar in blue, where \Zsolar is a fiducial solar metal mass fraction, equivalent to 0.02 in absolute units. For the lower two figures, supernova energy is shown as a solid line, wind kinetic energy as a dashed line, wind thermal energy as a dotted line and the energy in ionising photons as a dot-dashed line. The line marked ``radiation'' is the total energy in photons emitted from the star above the ionisation energy of hydrogen. See section \ref{methods:model} for a full description of the values used.}
  \label{methods:massloss_energy}
\end{figure}

In our simulations we simulate a single 15 \Msolar star in a variety of environments. The stellar wind model implemented in this paper is taken from the Padova stellar evolution models \citep{Marigo:2008p1413}. The initial velocity of the wind is set to the escape velocity of the star. \cite{Kudritzki2000} find that wind velocities only noticeably exceed the escape velocity for star more massive than the one modelled in this paper. The temperature of the gas ejected is taken to be the surface temperature of the star. While \cite{Runacres2005a} argue that the temperature decreases rapidly once it leaves the star, the kinetic energy of the wind dominates by roughly three orders of magnitude and thus the precise temperature of the wind is unimportant. For the metallicity of the wind, we assume a surface metallicity for our star equal to that of the external medium for all simulations. 
The lifetime of the star is allowed to vary with metallicity as per the \cite{Marigo:2008p1413} model. Based on the same model, the lifetime of the star is set to 13.2 Myr for a star of \Zsolar and 15.8 Myr for a star of 0.1 \Zsolar, where \Zsolar is the solar metal mass fraction, set to a fiducial value of 0.02 in absolute units. For runs including the radiative transfer of ionising photons, we produce a set of metallicity and age-dependent spectra for a 15 \Msolar star using the Starburst99 web-based software and data package \citep{Leitherer:1999p2023}. Once the star has reached the end of its lifetime, it explodes as a supernova. We use a supernova energy of $1.2\times10^{51}$ergs and a remnant mass of 1.5 \Msolar as per \cite{Kovetz:2009p1831,Smartt:2009p1830}. A metallicity of 6.5 \Zsolar is used for the gas ejected by the supernova explosion, which we derive from the results of \cite{Chieffi2013}.
Values for cumulative energy and mass input from the star as winds, ionising photons and supernova explosions are given in Figure \ref{methods:massloss_energy}. It is worth noting that the total energy emitted in ionising photons exceeds the supernova energy. However, it is not guaranteed that all of this energy will couple with the surrounding gas as a kinetic shock. The energy in ionising photons from the 0.1 \Zsolar star is higher than that of the \Zsolar star owing to the lower opacity of the former. The energy in ionising photons decreases slowly over the lifetime of the star as it expands and its surface temperature drops. The total energy available from stellar winds for this star is roughly three orders of magnitude lower than the supernova energy. Winds from more metal-rich stars eject proportionally more mass at higher velocities than more metal-poor stars, again due to the higher opacity that allows greater coupling of photons to the surface ions of the star.

We run this stellar model in a variety of uniform media with different initial densities and metallicities. Details of these simulations are given section \ref{methods:numsim}. The star is positioned at the centre of the simulation volume, and is static with respect to the external medium. The wind is imposed on the grid by incrementing the density, momentum and thermal energy of the grid cells inside a sphere of radius 20 cells at the highest refinement level with an inverse-square distribution to give a consistent mass in each spherical shell. We use a sphere to impose the solution rather than a single point in order to attempt to minimise the effect of grid artefacts and produce a spherically-shaped wind (note also that momentum cannot be deposited onto the grid in a spherical configuration in a single cell, since the velocity vectors would cancel each other). For the weak wind early in the star's lifetime we find that some grid artefacts are unavoidable. This is discussed further in section \ref{evolution:overview}. The supernova energy and mass is deposited as a thermal pulse at the centre of the grid. Photons, likewise, are deposited onto the centre of the grid and evolve subsequently according to the prescription 
described in \cite{Rosdahl2013}.

\subsection{Numerical simulations}
\label{methods:numsim}

\begin{table*}
\begin{tabular}{l c c c c c c c c c}
   \textbf{Name} & \textbf{n$_{H,ini}$ / \atcc} & \textbf{T$_{ini}$ / K} & \textbf{Z$_{ini}$ / \Zsolar} & \textbf{SNe?} & \textbf{Winds?} & \textbf{RHD?} \\
  \hline
 \simname{N0.1ZsoS} & 0.1 & 62 & 1.0 & \tick & & \\
 \simname{N0.1ZsoSW} & 0.1 & 62 & 1.0 & \tick & \tick & \\
 \simname{N0.1ZsoSR} & 0.1 & 62 & 1.0 & \tick &  & \tick  \\
 \simname{N0.1ZsoSWR} & 0.1 & 62 & 1.0 & \tick & \tick & \tick \\
 \simname{N0.1ZloSWR} & 0.1 & 94 & 0.1 & \tick & \tick & \tick \\
  \hline
 \simname{N0.5ZsoSWR} & 0.5 & 31 & 1.0 & \tick & \tick & \tick \\
 \simname{N0.5ZloSWR} & 0.5 & 75 & 0.1 & \tick & \tick & \tick \\
  \hline
 \simname{N5ZsoSWR} & 5 & 12 & 1.0 & \tick & \tick & \tick \\
 \simname{N5ZloSWR} & 5 & 32 & 0.1 & \tick & \tick & \tick \\
  \hline
 \simname{N30ZsoS} & 30 & 8.2 & 1.0 & \tick & & & \\
 \simname{N30ZsoSW} & 30 & 8.2 & 1.0 & \tick & \tick & \\
 \simname{N30ZsoSR} & 30 & 8.2 & 1.0 & \tick &  & \tick & \\
 \simname{N30ZsoSWR} & 30 & 8.2 & 1.0 & \tick & \tick & \tick \\
 \simname{N30ZloSWR} & 30 & 13 & 0.1 & \tick & \tick & \tick \\
  \hline
 \simname{N100ZsoSWR} & 100 & 8.2 & 1.0 & \tick & \tick & \tick \\
 \simname{N100ZloSWR} & 100 & 9.9 & 0.1 & \tick & \tick & \tick \\
  \hline
\end{tabular}
  \caption{Table of properties of numerical simulations included in this paper. n$_{H,ini}$, T$_{ini}$ and Z$_{ini}$ refer to the initial hydrogen number density, initial temperature and initial metallicity of the simulation volume around the star (we determine T$_{ini}$ by allowing a low-resolution volume with the same density and metallicity to relax to a given temperature, and then set the initial temperature of the simulation to this value). ``\simname{N}'' refers to the initial hydrogen number density, given by the number after it. ``\simname{Z}'' denotes the initial metallicity, which is either ``solar'' (Z=\Zsolar) given by ``\simname{so}'', or ``low'' (Z=0.1 \Zsolar) given by ``\simname{lo}''. Letters ``\simname{S}'', ``\simname{W}'' and ``\simname{R}'' denote that a supernova (SNe), stellar winds and radiation hydrodynamics (RHD) respectively are included in the simulation. 
See section \ref{methods:numsim} for full details of the simulations run.}
\label{methods:numsimtable} 
\end{table*}

We run our simulations using \textsc{RAMSES-RT} \citep{Rosdahl2013}, a radiation-hydrodynamics (RHD) extension of the AMR code \textsc{RAMSES} \citep{Teyssier:2002p533}, which includes the propagation of photons and their on-the-fly interaction with gas via photoionisation and heating of hydrogen and helium. The advection of photons between grid cells is described with a first order moment method and the set of radiation transport equations is closed with the M1 relation for the Eddington tensor. 
RAMSES-RT solves the non-equilibrium evolution of the ionisation fractions of hydrogen and helium, along with ionising photon fluxes and the gas temperature in each grid cell. Metal cooling is added assuming photoionisation equilibrium with a \cite{Haardt:1996p1457} redshift 0 UV background. The spectrum of ionising photons from the star is modelled by three photon groups, bracketed by the ionisation energies of HI, HeI, and HeII. We ignore photons at sub-ionising energies, and we ignore radiation-dust interactions. As stated in the introduction, we also ignore radiation pressure. Rosdahl et al (2015, in prep.) discusses direct radiation pressure from photoionisation in similar simulation setups and find it to be negligible compared to the effect of photoionisation heating.

The simulations we ran are listed in Table \ref{methods:numsimtable}. In the runs where the medium around the star has a number density of 0.1 \atcc and 30 \atcc at solar metallicity, we perform a series of experiments in which different stellar processes are included (namely stellar winds, photoionisation and supernova feedback), in order to investigate the relative impact of each feedback process in isolation. These densities are selected as they represent roughly the densities found in the diffuse and dense phases of the \ISM. 0.1 \atcc is also a point of comparison for previous works. We then run a grid of simulations with every physical process included at five different densities around these values, at both solar and 10\% of solar metallicity. For each simulation we run a low resolution equivalent at the given initial density and metallicity and allow the temperature to relax to an equilibrium value. We then set the initial temperature to this value. To simplify the model and allow us to study the impact of the model star in a controlled environment, we do not include external sources of heating or turbulence. This will be explored in future works, though \cite{Raga2012} and \cite{Tremblin2014} produce 1D models for HII regions in the presence of turbulence. Using \cite{Cioffi1988} we calculate that the supernova remnant will merge with the \ISM due to turbulence on timescales of around 20 Myr at 100 \atcc and 70 Myr at 0.1 \atcc, several times longer than the time over which we follow the supernova remnant. We do not consider self-gravity in this work, since the structures produced in this work are more or less spherically symmetric and as such are not strongly self-gravitating. In a more realistic medium, the external density field will dominate the gravitational field in the \CSM and subsequent supernova remnant.

Each simulation is run in a cubic box with length 4.8kpc and a root grid with $64^3$ cells. The large box size ws originally chosen in order to limit artefacts arising from the Poisson solver in \textsc{Ramses}, which uses periodic boundary conditions, though we found that self-gravity had a limited effect on the results and did not include it in the final runs. We then allow the simulations to refine up to a maximum spatial resolution of 0.6 pc, 2$^{-13}$ of the box length. A cell is allowed to refine if the fractional difference in either pressure or density with a neighbouring cell exceeds 0.2 (though tests conclude that results do not vary significantly if this threshold is varied). We calibrate the choice of maximum spatial resolution based on tests of the \simname{N0.1ZsoSW}, \simname{N30ZsoSW} and \simname{N30ZsoSR} simulations. The spatial resolution was selected such that the evolution of energy and shock radius in the simulation was unchanged to within a few percent by a factor 2 decrease in grid cell size. The only exception is the \simname{N30ZsoS} simulation, which has a short cooling time for the spatial resolution. Here, we use an extra level of refinement, corresponding to 0.3 pc maximum spatial resolution, although the difference between this and a run taken at the default resolution is small.


\section{Evolution of the CSM}
\label{evolution}

\begin{figure}
\centerline{\includegraphics[width=1.0\hsize]{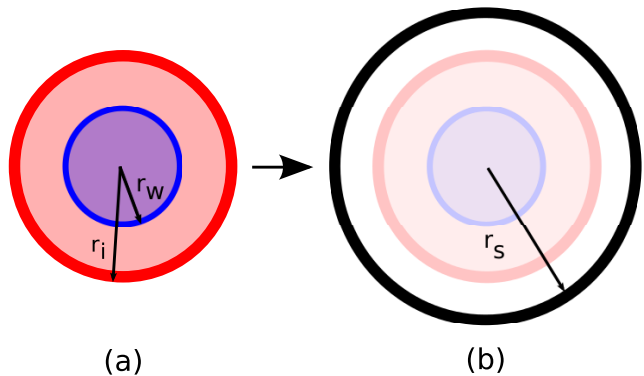}}
  \caption{A schematic drawing of the structures around the star before and after the supernova event. Each process creates a heated underdensity surrounded by an overdense shell. Left: the \protect\CSM just before the supernova. The ionised gas forms a shell to radius r$_{i}$. Inside r$_{i}$ is the wind-blown cavity with its own (typically weak) shell at r$_w$. Around the star is a small free-streaming radius, where the wind has yet to shock against the \protect\CSM gas (c.f. \protect\cite{Weaver1977}). Right: the \protect\CSM after the supernova. The supernova sweeps up and destroys the existing structures in the \protect\CSM and creates its own hot, diffuse bubble and a dense shell of radius r$_s$ that spreads into the \protect\ISM (pre-existing structures overlaid for comparison).}
  \label{introduction:cartoon}
\end{figure}

\subsection{Overview}
\label{evolution:overview}

\begin{figure}
\centerline{\includegraphics[width=0.98\hsize]{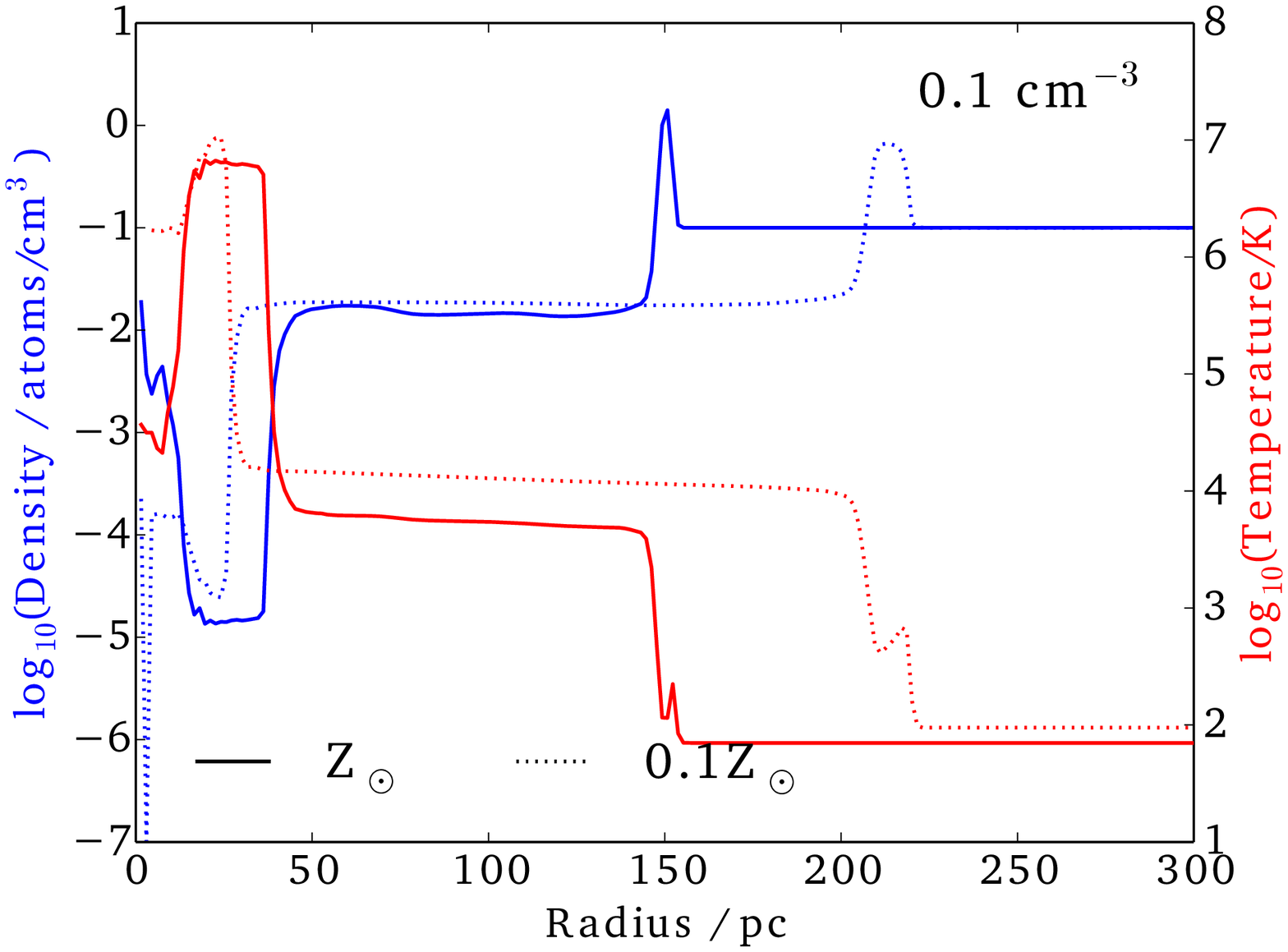}}
\centerline{\includegraphics[width=0.98\hsize]{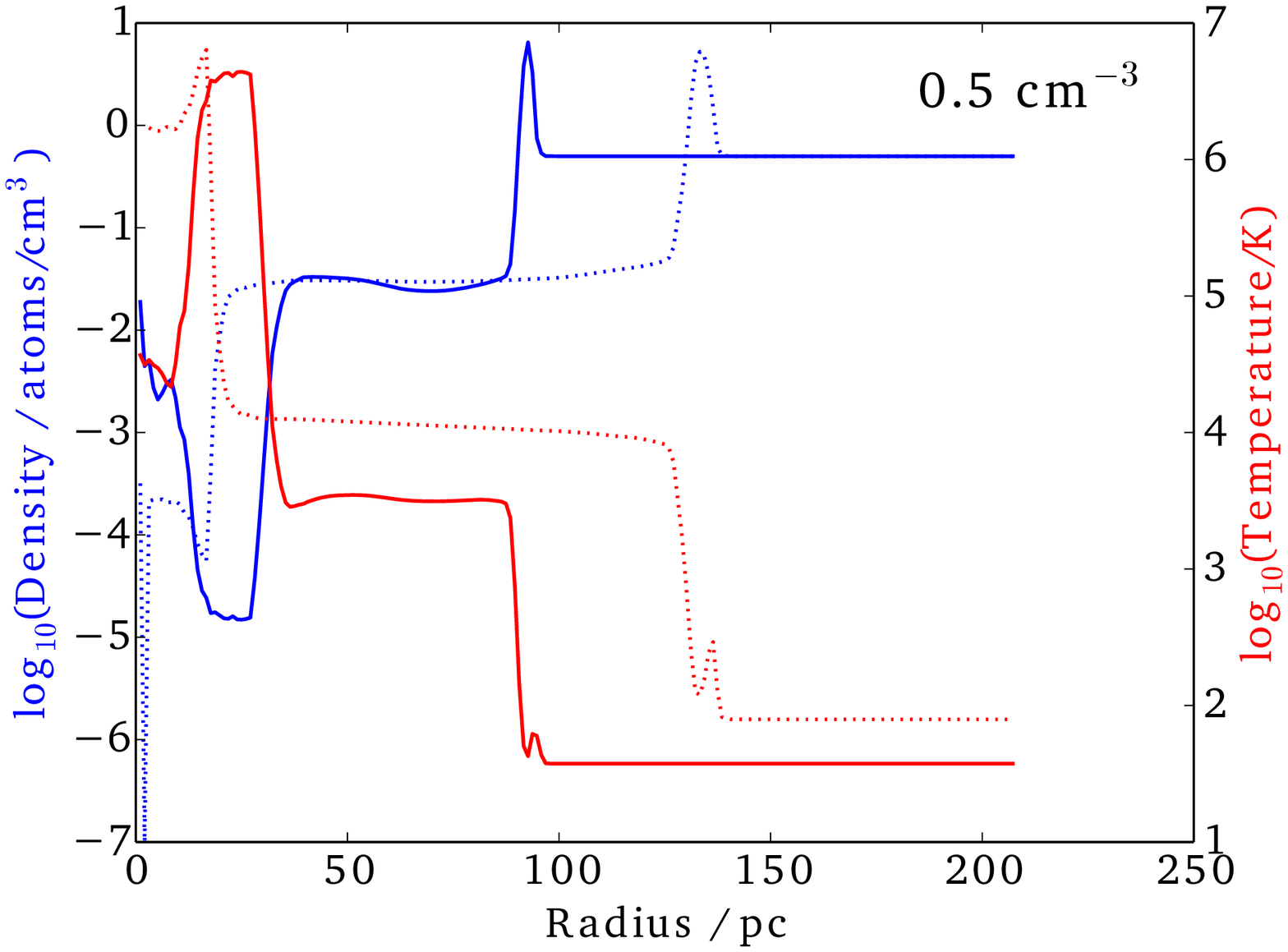}}
\centerline{\includegraphics[width=0.98\hsize]{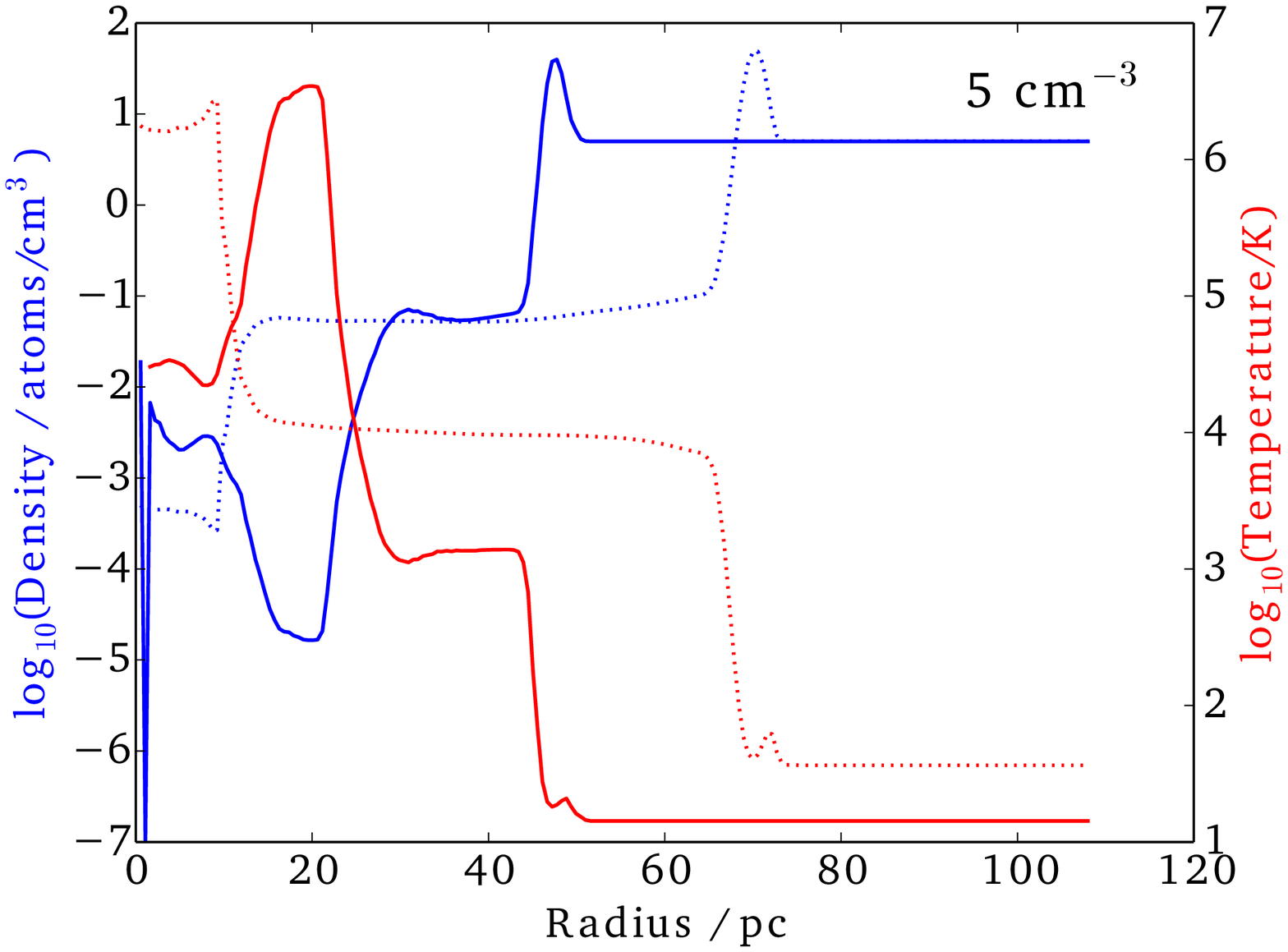}}
  \caption{Density and temperature radial profiles at t$_{SN}$  plotted for each of the runs containing both winds and photoionisation, comparing solar to 10\% solar metallicity simulations. Density is shown in blue and temperature in red.  A solid line indicates solar metallicity and a dotted line 0.1 \Zsolar. \simname{N0.1ZsoSWR} and \simname{N0.1ZloSWR} are plotted on the top row, \simname{N0.5ZsoSWR} and \simname{N0.5ZloSWR} on the middle row and \simname{N5ZsoSWR} and \simname{N5ZloSWR} on the bottom row. The other runs are plotted in Figure \ref{evolution:basic_profiles_dens}. The value of t$_{SN}$ used depends on the metallicity of the star in the simulation, as given in section \protect\ref{methods:model}. The value in each radial bin is found by averaging the values for all points inside that bin.}
  \label{evolution:basic_profiles}
\end{figure}

\begin{figure}
\centerline{\includegraphics[width=0.98\hsize]{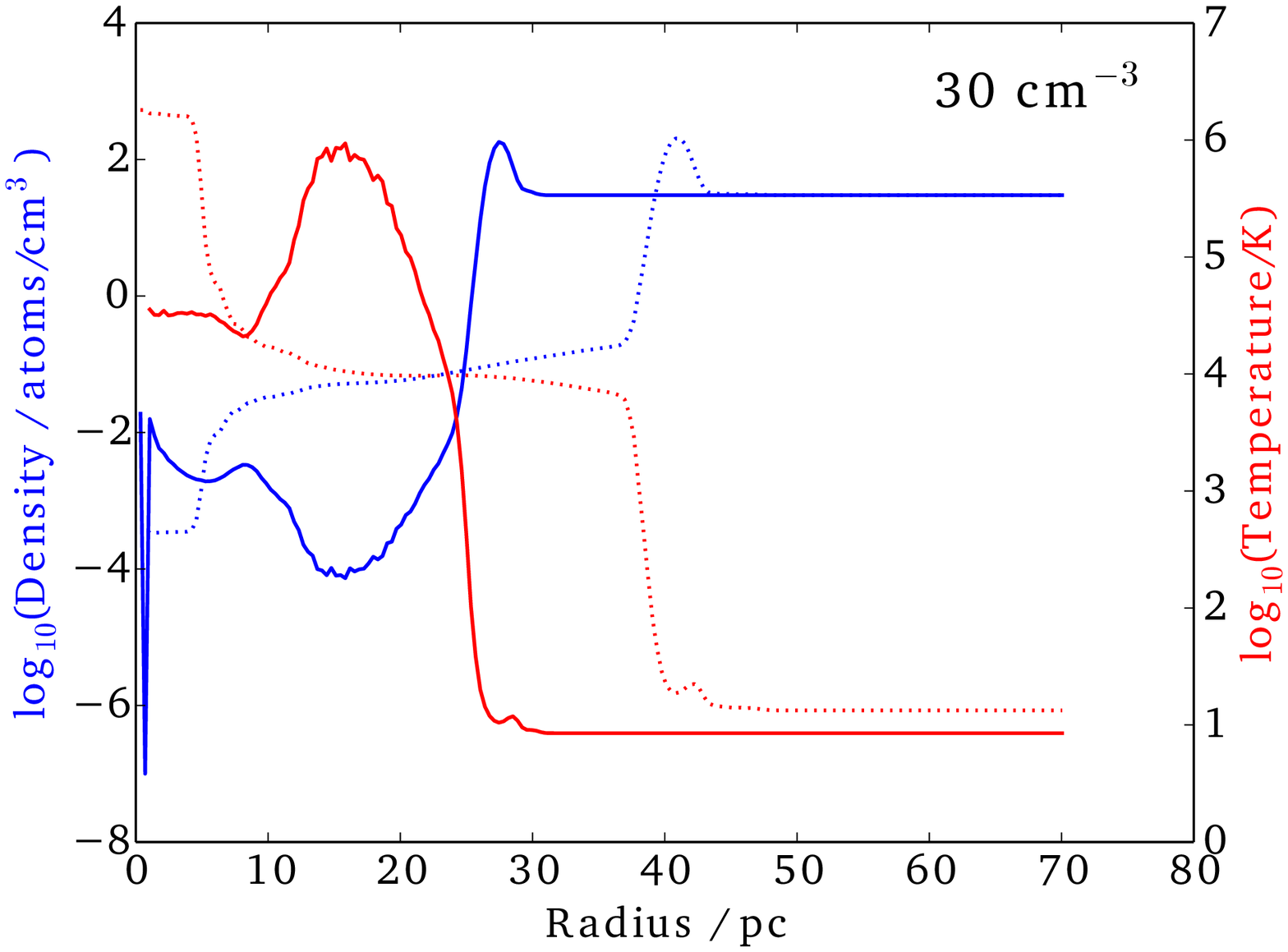}}
\centerline{\includegraphics[width=0.98\hsize]{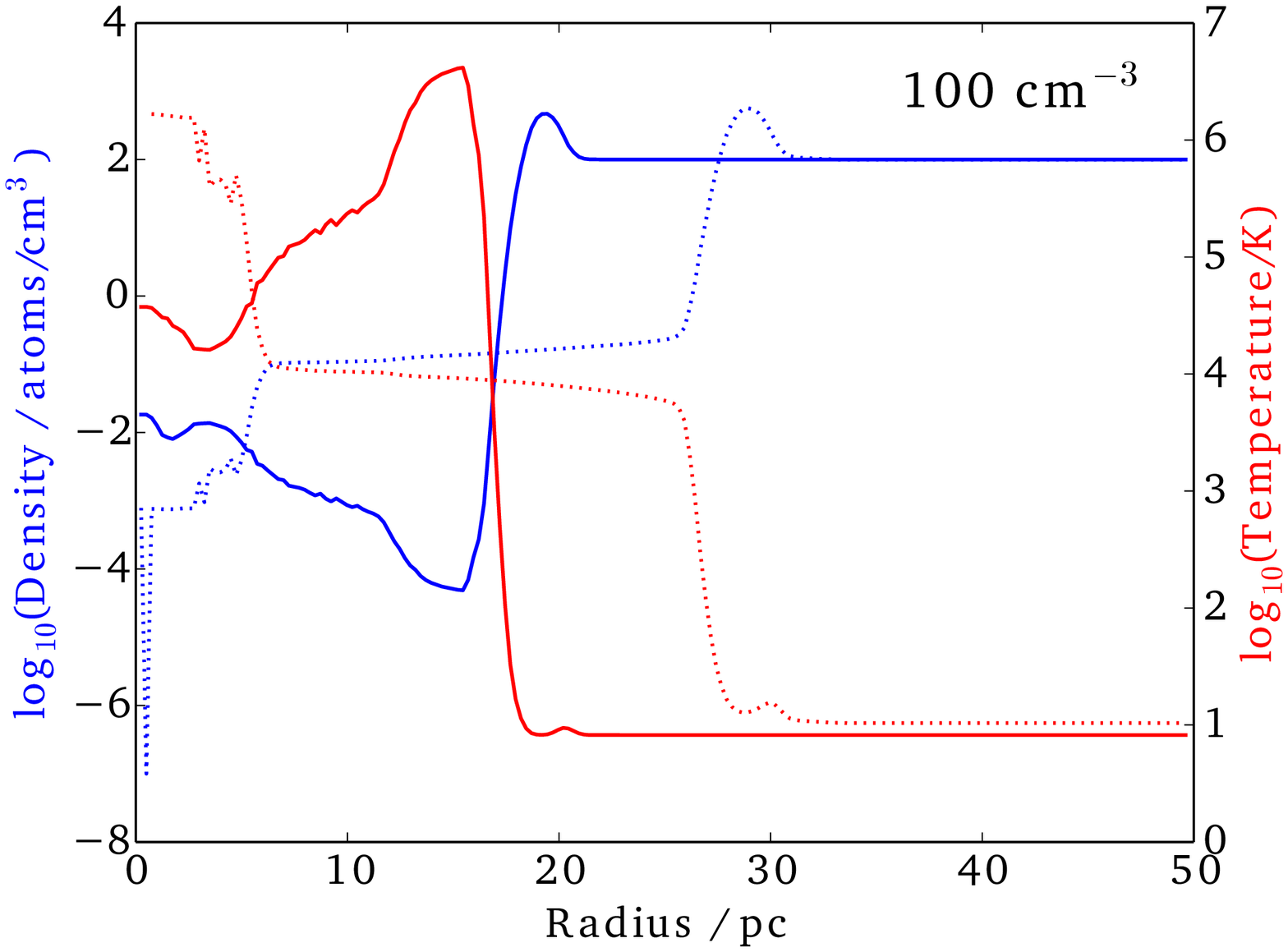}}
  \caption{As in Figure \protect\ref{evolution:basic_profiles}, but for the runs \simname{N30ZsoSWR} and \simname{N30ZloSWR} (top), and \simname{N100ZsoSWR} and \simname{N100ZloSWR} (bottom).}
  \label{evolution:basic_profiles_dens}
\end{figure}

\begin{figure}
\centerline{\includegraphics[width=1.0\hsize]{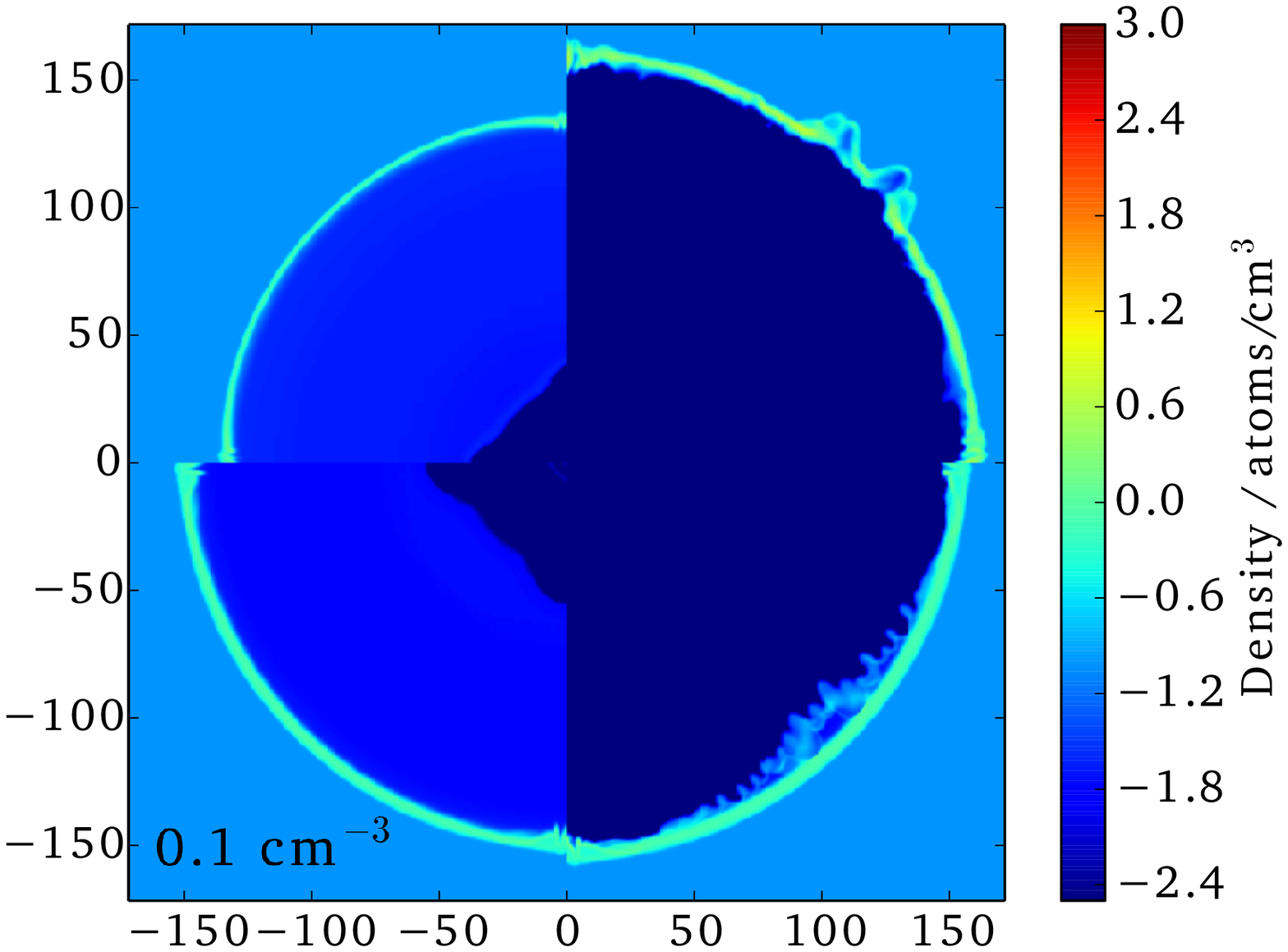}}
\centerline{\includegraphics[width=1.0\hsize]{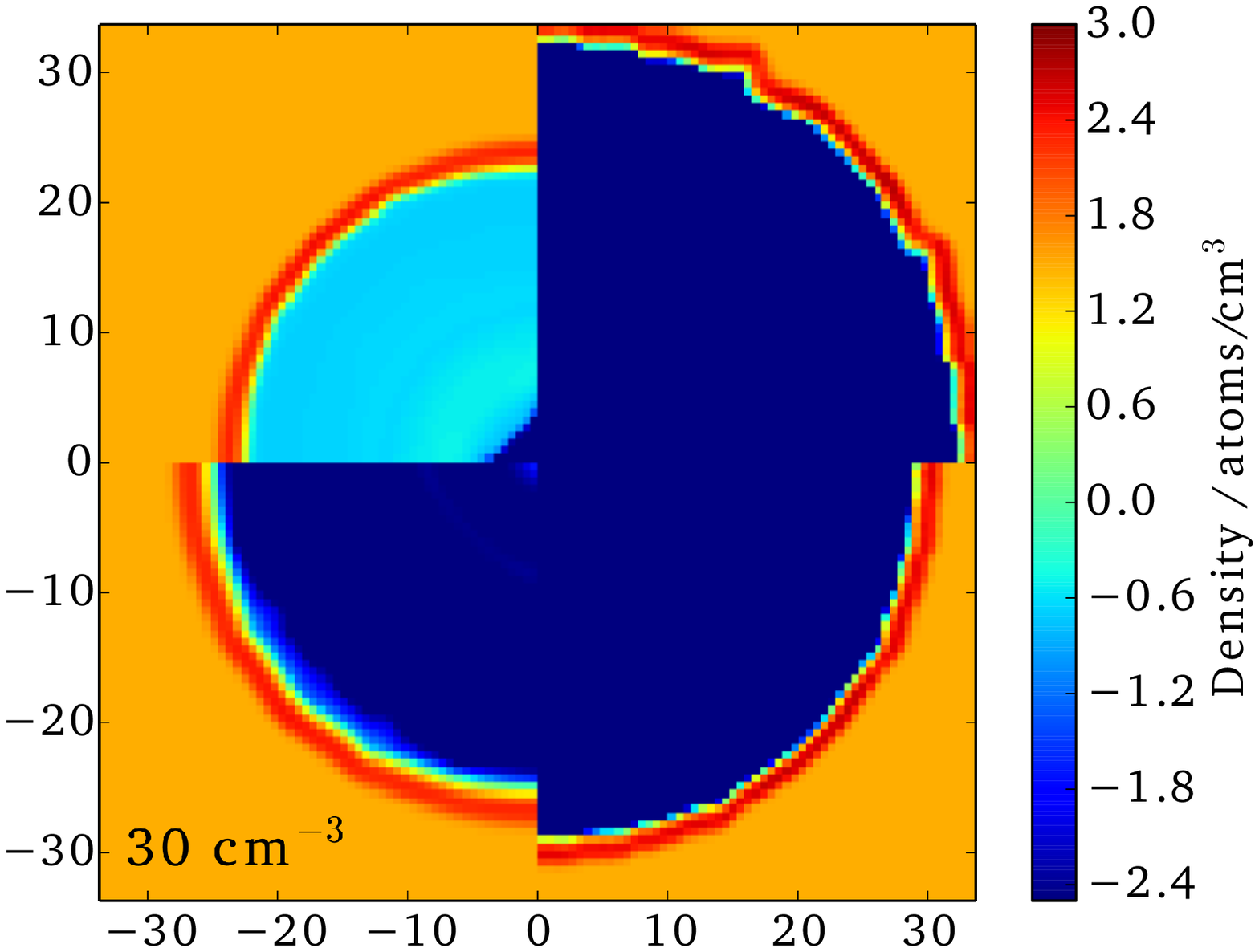}}
  \caption{Slices in density through the simulation volume in the plane of the star and oriented along the grid z-axis for runs \simname{N0.1ZsoSWR} (top) and \simname{N30ZsoSWR} (bottom). The quadrants show the simulation at times, arranged counter-clockwise from top left, 10Myr, t$_{SN}$, 1Myr after t$_{SN}$ and 2Myr after t$_{SN}$ (as in Figure \protect\ref{evolution:basic_profiles}), where t$_{SN}$ is 13.1Myr. The image axes are in parsecs.}
  \label{evolution:slices}
\end{figure}
%
%
%

The structures produced by the star in the \CSM are broadly similar for all runs, though the precise properties vary for each run. A schematic view is given in Figure \ref{introduction:cartoon}. As the star evolves, an ionisation front at $r_i$ expands outwards, bounded by a dense shell of swept-up material, with a wind-driven bubble at $r_w$ embedded inside it. Then, once the star explodes as a supernova, a supernova-driven shock propagates outwards at $r_s$, erasing the previous structures, interacting with the shell of the ionisation front and propagating into the unshocked \ISM. Radial profiles for each run containing both stellar winds and radiation hydrodynamics are shown in Figures \ref{evolution:basic_profiles} and \ref{evolution:basic_profiles_dens}. The gas inside the ionisation front ($r < r_i$) reaches a temperature of around $10^4$ K inside a sphere of radius $r_i$, where $r_i$ is the radius of the ionisation front (neglecting the thickness of the shell at $r_i$). The precise temperature found in observed HII regions varies between 5000 K and 15000 K as a function of gas density and metallicity \citep{Draine2011}, but for the purposes of this work we use $10^4$ K in our analysis since it matches our solar metallicity results well. At first the gas expands hydrostatically to the Str\"omgren radius, the radius at which the number of recombinations equals the number of photoionisations. The pressure difference between the ionised gas and its surroundings causes the gas to expand outwards into the neutral \ISM. As it does so, it creates an overdense shell at $r_i$, gathering matter from the external medium as well as matter driven outwards by the shock as it attempts to regain pressure equilibrium with its surroundings. As the star evolves, its ionising luminosity decreases as the star expands and its surface temperature drops. In the solar metallicity case, this causes the ionisation fraction of the photoionised gas (i.e. the fraction of atoms that have been photoionised) to drop as rate of recombination events rises above the rate of photoionisation events. As a result, the temperature of the photoionised gas drops. For the runs at 0.1 \Zsolar, the luminosity in ionising photons is much higher. As a result, the ionised gas remains at roughly $10^4$ K. In addition, the radius of the ionisation front is larger by around 50\%. The density of the ionised gas, however, remains the same, since the rate of expansion of the shell is limited by the sound speed in the ionised gas, which is approximately 10 km/s at $10^4$ K. The larger radius can be attributed instead to a larger initial Str\"omgren sphere, as discussed in section \ref{evolution:photo}.

Inside the ionisation front is a wind-driven bubble of radius $r_w$. A free-streaming wind surrounds the star, as matter at the surface temperature of the star flows outwards. This material eventually shocks against the \CSM, heating the gas to $10^6 - 10^7$ K. Since the energy in the wind is much lower than the energy in ionising photons, the pressure difference created by the wind is lower than that created by the ionising photons, and as such $r_w$ typically lags behind $r_i$. There is a weak overdensity around the wind bubble, but most of this matter is photoheated and swept up by the photoionised bubble. In the runs at 0.1 \Zsolar, the wind is weaker still due to the star's reduced opacity, meaning fewer particles are expelled from the surface of the star. Similar structures to those in the \Zsolar runs are seen in these simulations, with comparable temperatures inside the wind bubble but with less mass redistributed by the wind. At 0.1 \Zsolar, the free-streaming wind phase is barely apparent. For the densest cases (Figure \ref{evolution:basic_profiles_dens}), the wind bubble catches up to the ionisation front. The consequence of this is that the outer edge of the bubble is heated to $10^6$ K, whereas the unshocked wind inside the bubble is heated to $10^4$ K by photons, leading to what might appear to be a smaller HII region embedded within a wind bubble. The interaction between the HII region and the wind is discussed further in section \ref{evolution:winds}.

\rev{Although the simulation is performed in a uniform medium, instabilities develop on the surface of the wind bubble. These can be observed in Figure \ref{evolution:slices}. The greatest deviations from spherical symmetry are aligned with the grid. In the absence of external turbulence, the most significant seed for instabilities is the grid structure itself. \cite{Ntormousi2011} note that along the grid axes, the cells are spaced closer together than cells along a diagonal. This means that the fluid is better resolved for surfaces normal to the grid axes. In these directions, the shell can be more easily compressed and is more susceptible to instabilities such as those described by \citep{Vishniac1983}. This is an issue for our simulations, in which the winds are weak and the flows are marginally more efficient along the grid axes, where there is a higher effective resolution. \cite{Ntormousi2011} note that increased resolution does not help reduce the grid-aligned instabilities as the grid-aligned and grid-diagonal issue remains, and that the length scales required to achieve convergence cannot be reached with the available computational resources. This is because thermal instabilities are governed by the Field length \citep{Koyama2004}, which at the dense shell is much smaller than the maximum spatial resolution achievable by contemporary 3D simulations of stellar feedback. Despite these issues, our results have converged with spatial resolution, as stated in section \ref{methods:numsim}.}

\rev{Prior to the supernova, the ionisation front is largely spherical, though some fluctuations can be observed in the shell (again, largely in the direction of the grid axes). Once the supernova occurs, the shock passes through the existing structures, gaining structure from the asphericity of the wind bubble, and causing fluctuations in the shell of the ionisation front (which is now the shell of the supernova remnant). This effect is most apparent in the \simname{N0.1ZsoSWR} image. In the \simname{N30ZsoSWR} run, the wind has already reached the shell of the ionisation front. We discuss in more detail in section \ref{evolution:winds}.}

\subsection{Evolution of the Ionisation Front}
\label{evolution:photo}

\begin{figure}
\centerline{\includegraphics[width=0.99\hsize]{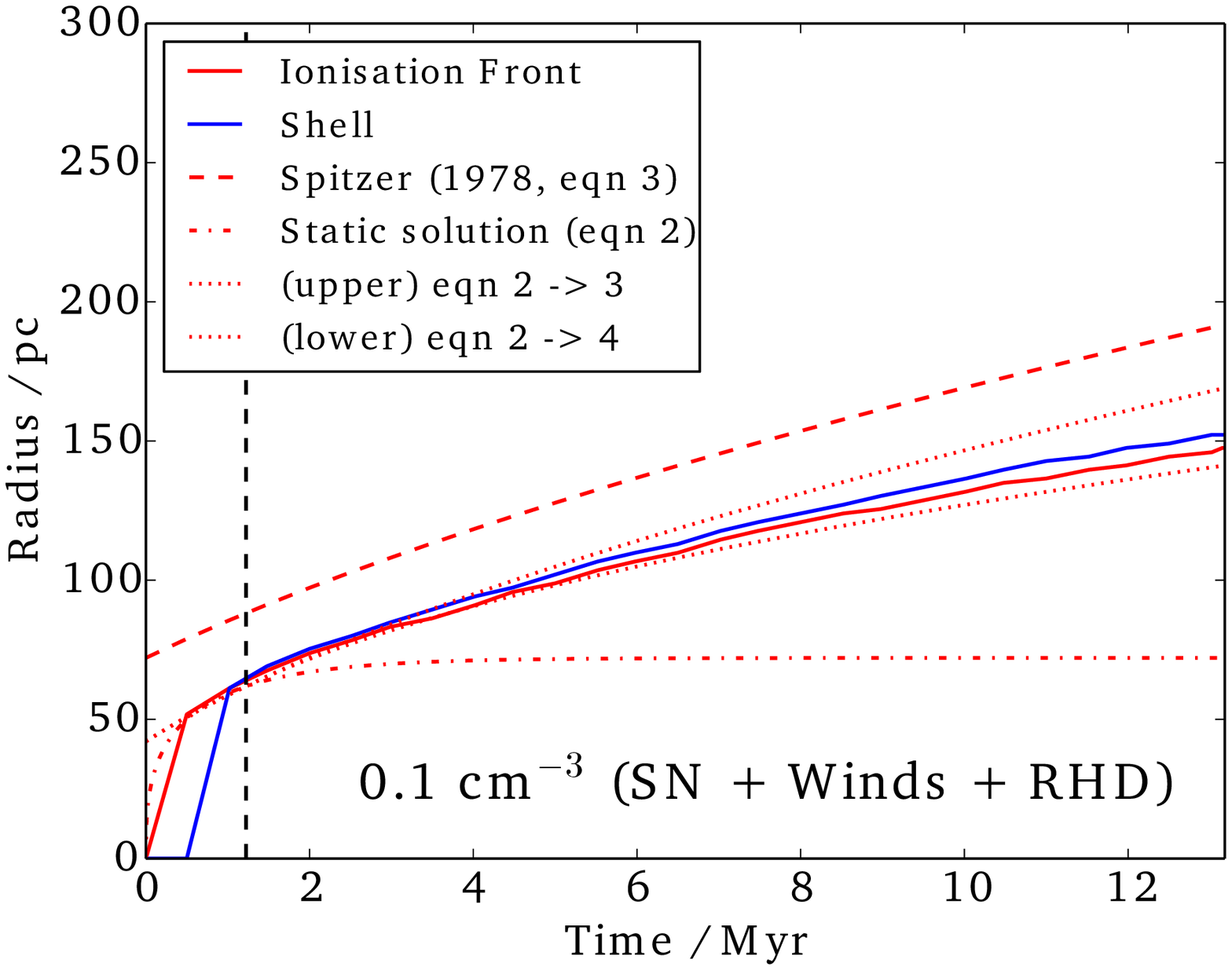}}
\centerline{\includegraphics[width=0.99\hsize]{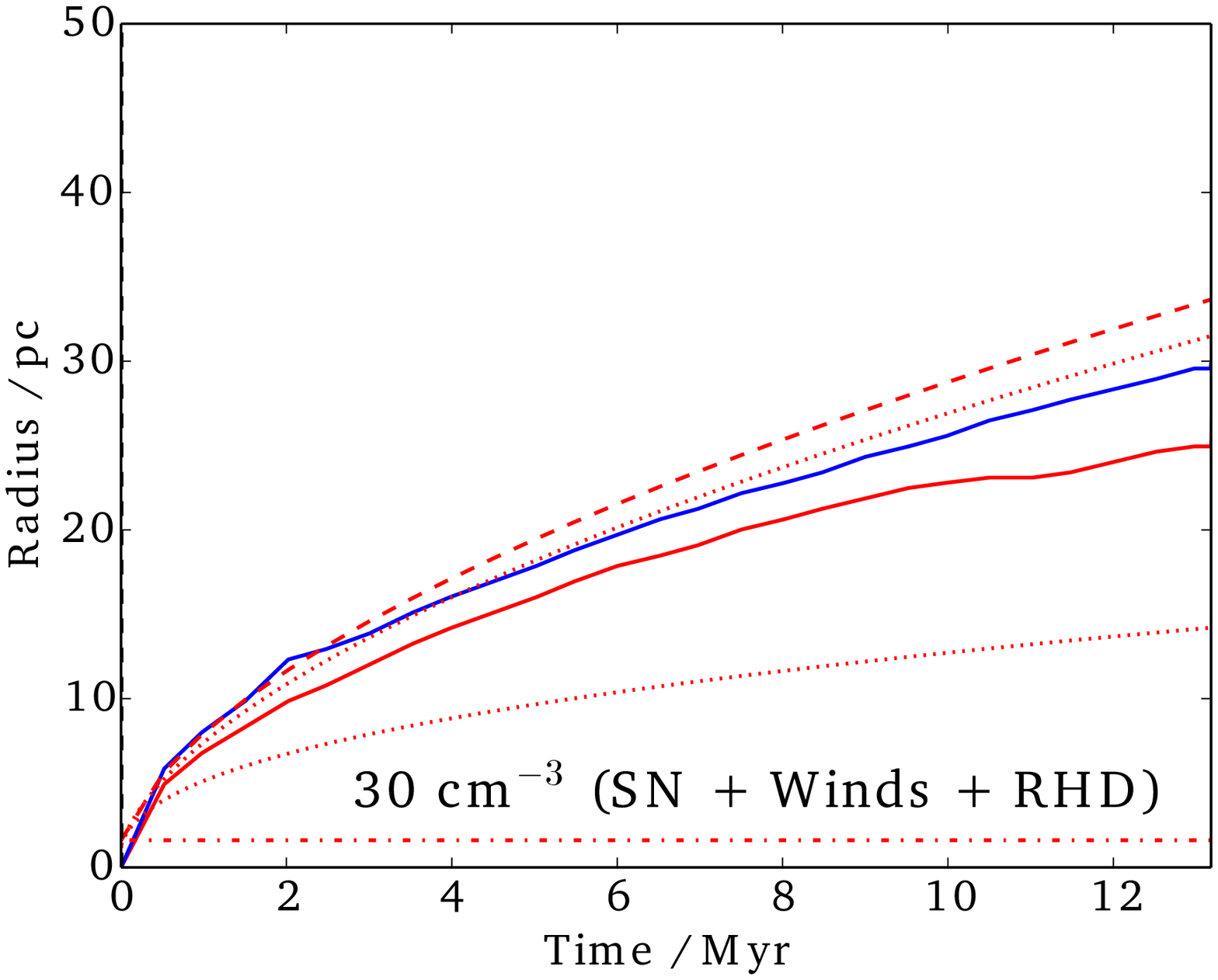}}
  \caption{Radius of the ionisation front against time up to t$_{SN}$. The upper Figure shows data for the \simname{N0.1ZsoSWR} run, while the lower Figure shows the \simname{N30ZsoSWR} run data. The solid blue line is the extent of the dense shell (the maximum radius at which $n_H > n_{H,ini}$), while the solid red line is the maximum radius at which more than 10\% of hydrogen atoms are ionised.  A vertical dashed line is plotted at the recombination time at the given density, $t_{rec}$. The upper dashed curve is the \protect\cite{SpitzerLyman1978} given by equation \ref{evolution:expansion_wave_spitzer}. The upper dotted curve is the same equation, but solved starting from time $t_{rec}$ using the radius equation \ref{evolution:expansion_stromgren} at $t_{rec}$. The bottom curve uses the same method but uses equation \ref{evolution:expansion_wave_shell} instead. This equation assumes that the falling UV photon flux leads to the supply of ionised gas being held constant. The dot-dashed curve shows the hydrostatic evolution to the Str\"omgren radius $r_{st}$ given in equation \ref{evolution:expansion_stromgren}.}
  \label{evolution:ionradii}
\end{figure}

The expansion of the ionisation front is characterised by two phases. The first occurs on the order of the recombination time, $t_{rec}$, where the ionisation front approaches the Str\"omgren radius $r_{st}$ \footnote{Note that this is distinct from the radius of the supernova remnant, which we label $r_{s}$}. This is the radius inside which the rate of recombination events between free electrons and ions is equal to the flux of ionising photons, and is given by:
\begin{equation}
  r_{st} = \left ( \frac{3}{4 \pi} \frac{S_*}{n_i n_e \alpha_B} \right )^\frac{1}{3}
  \label{evolution:stromgren_radius}
\end{equation}

where $S_*$ is the flux of ionising photons from the star in photons per unit time, $\alpha_B$ is the total recombination rate, and $n_i$ and $n_e$ are the ion and electron number density respectively. $n_e = n_i$ if the ionisation fraction $x = 1$, and hence for a fully ionised medium, $r_{st}^3 \propto n_i^{-2}$, which is an important result that will be referred to later in the paper. Note that this requires either a pure hydrogen \CSM or one in which helium is only singly ionised. Indeed, we find in our results that the HeIII fraction is negligible. $r_i$ reaches $r_{st}$, assuming no hydrodynamic response from the gas, according to

\begin{equation}
  r_i(t) = r_{st} \left (1 - e^{-n_i \alpha_B t} \right ).
  \label{evolution:expansion_stromgren}
\end{equation}

The second phase is the hydrodynamic response of the gas due to thermalisation to $10^4$ K by photoionisation. This phase is described analytically by \cite{SpitzerLyman1978}:

\begin{equation}
  r_i = r_{st} \left (1 + \frac{7}{4}\frac{C_i t}{r_{st}} \right )^\frac{4}{7}
  \label{evolution:expansion_wave_spitzer}
\end{equation}

where $C_i$ is the speed of sound in the ionised gas ($\simeq$ 10 km/s). In this phase, the photoionised gas is heated to approximately $10^4K$, which creates a pressure gradient at the ionisation front. This causes the density inside the ionised gas to drop and the remaining mass to be deposited around the ionisation front as a dense shell. The recombination time is inversely proportional to the density of the medium: 1.2 Myr for 0.1 \atcc and 4 kyr for 30 \atcc. Thus, for dense media, the ionisation front reaches $r_{st}$ on a timescale much shorter than the lifetime of the star. By contrast, the photons in more diffuse media take around 10\% of the lifetime of the star to reach $r_{st}$.  This has consequences for the evolution of the front. In Figure \ref{evolution:ionradii}, the ionisation front of the star in run \simname{N30ZsoSWR} reaches the \Stromgren~radius almost immediately, and then follows the Spitzer solution until the flux of ionising photons drops and the ionisation front expands less rapidly. By contrast, the 
ionisation front in run \simname{N0.1ZsoSWR} does not reach $r_{st}$ before it begins to respond hydrodynamically. However, the solutions are only weakly coupled: since the \Stromgren~radius is 70pc at 0.1 \atcc and 1.6pc at 30 \atcc, and the speed of sound for both densities inside the ionisation front is around 10 km/s $\simeq$ 10 pc/Myr, a sound wave would take five times as long to cross $r_{st}$ as the recombination time at 0.1 \atcc  (37 times at 30 \atcc). We thus introduce a solution in which the ionisation front expands to $r_{st}$, and then is allowed to expand according to equation \ref{evolution:expansion_wave_spitzer}. This solution is valid until the photon flux begins to fall significantly, and the ionisation fraction falls below 1. At this point equation \ref{evolution:stromgren_radius} is no longer applicable, since the number of recombinations per second exceeds the flux of photons. In one scenario, the photons can no longer ionise new gas, and the mass of the bubble is constant, i.e. $r_i^3 n_i$ = constant. Solving the jump conditions given in \cite{SpitzerLyman1978} as used to derive equation \ref{evolution:expansion_wave_spitzer}, but with $r_i^3 \propto n_i^{-1}$ instead of $n_i^{-2}$, we find

\begin{equation}
  r_i = r_0 \left (1 + \frac{5}{2}\frac{C_i (t-t_{rec})}{r_0} \right )^\frac{2}{5}
  \label{evolution:expansion_wave_shell}
\end{equation}

where $r_0$ is the radius at $t_{rec}$ ($\leq r_{st}$). We also plot this in Figure \ref{evolution:ionradii}. Run \simname{N0.1ZsoSWR} follows the Spitzer solution closely before falling between it and equation \ref{evolution:expansion_wave_shell} after 4 Myr. While the gas is no longer being ionised up to $r_{st}$ as in the Spitzer solution, there is some residual photoionisation that keeps the gas partially photoionised.

The momentum and kinetic energy of the shell can be approximated with reasonable accuracy by using these radial solutions, and assuming that all the shell mass is concentrated at $r_i$, travelling at d${r_i}$/d$t$. The mass of the shell can be calculated by subtracting the ionised bubble mass, calculated using $r_s \propto n^{-2/3}$ as above, although the transition between fully-ionised and partially-ionised regimes as the star's ionising luminosity drops complicates finding an exact analytic solution. This solution agrees roughly with the energy and momentum of the shell given in \cite{Walch2012}. Despite the large quantity of energy in ionising photons leaving the star throughout its lifetime, only 0.1-0.01\% of this energy is transferred to kinetic energy in the shell, most of it being lost as radiation. The key impact that photoionisation has in terms of feedback from the star is to alter the density of the gas around the star prior to supernova. We return to this subject in section \ref{supernova}.

In our simulations we also include helium ionisation, which is provided by default in \textsc{Ramses-RT}. Typically, photoheating from hydrogen is the dominant process, and we do not notice much difference if we remove helium . Even before the temperature of the star has dropped noticeably, the helium inside the ionisation front is not completely ionised to HeII, and very little is ionised to HeIII. Many photons at energies that ionise helium to HeIII are able to escape the ionisation front entirely. A small amount of leakage, i.e. photons escaping the shell at $r_i$, is also observed in hydrogen-ionising photons in the runs at 0.1\atcc. Since the gas began responding hydrodynamically at this density before the ionisation front had reached $r_{st}$, the ionisation front lags behind $r_{st}$. As a result, a number of photons reach the shell and some are able to pass through it without being absorbed. Subsequently, the value of $r_0$, given by $r_i(t_{rec})$, that we use in equation \ref{evolution:expansion_wave_shell} is lower than $r_{st}$ for the run at 0.1\atcc. Additionally, the hot, shocked gas inside the wind bubble thermally ionises the \CSM up to $r_w$. This allows the photons to pass up to $r_w$ without being absorbed by neutral hydrogen.

We should note that the external medium in our simulations is static and largely unpressurised. \cite{Raga2012} introduce a term to Spitzer's equation to account for thermal and turbulent pressure in the \CSM, and determine that there is a point at which the pressure inside the ionisation front is equal to that outside, and the front cannot expand further. \cite{Tremblin2014} expand on this by simulating ionisation fronts with external turbulence. They find that while the solutions are constrained by external pressure, existing momentum in the shell can cause the simulated shells to overshoot \cite{Raga2012}'s analytic solution.

One further consideration is that metal cooling and heating rates in photoionised gas are typically different from those in neutral gas. \cite{Draine2011} states that the equilibrium temperature in the gas may vary from 5000K to 15000K, depending on its metallicity, density and the photon flux. In this work we do not include these rates, though in practice they may become important for modelling HII regions accurately.

\subsection{Expansion of the Wind Bubble}
\label{evolution:winds}

\begin{figure*}
\centerline{\includegraphics[width=0.49\hsize]{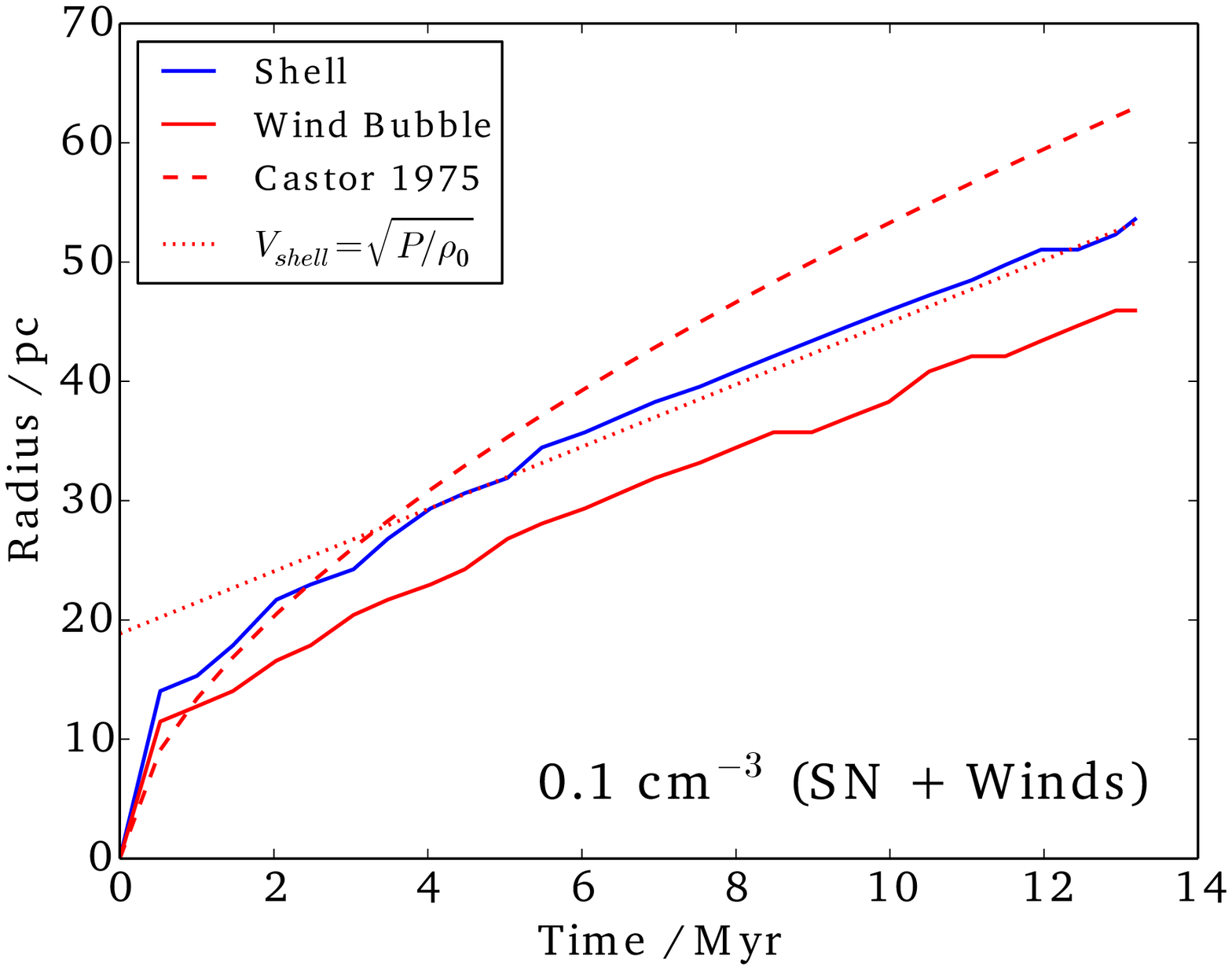}  \includegraphics[width=0.49\hsize]{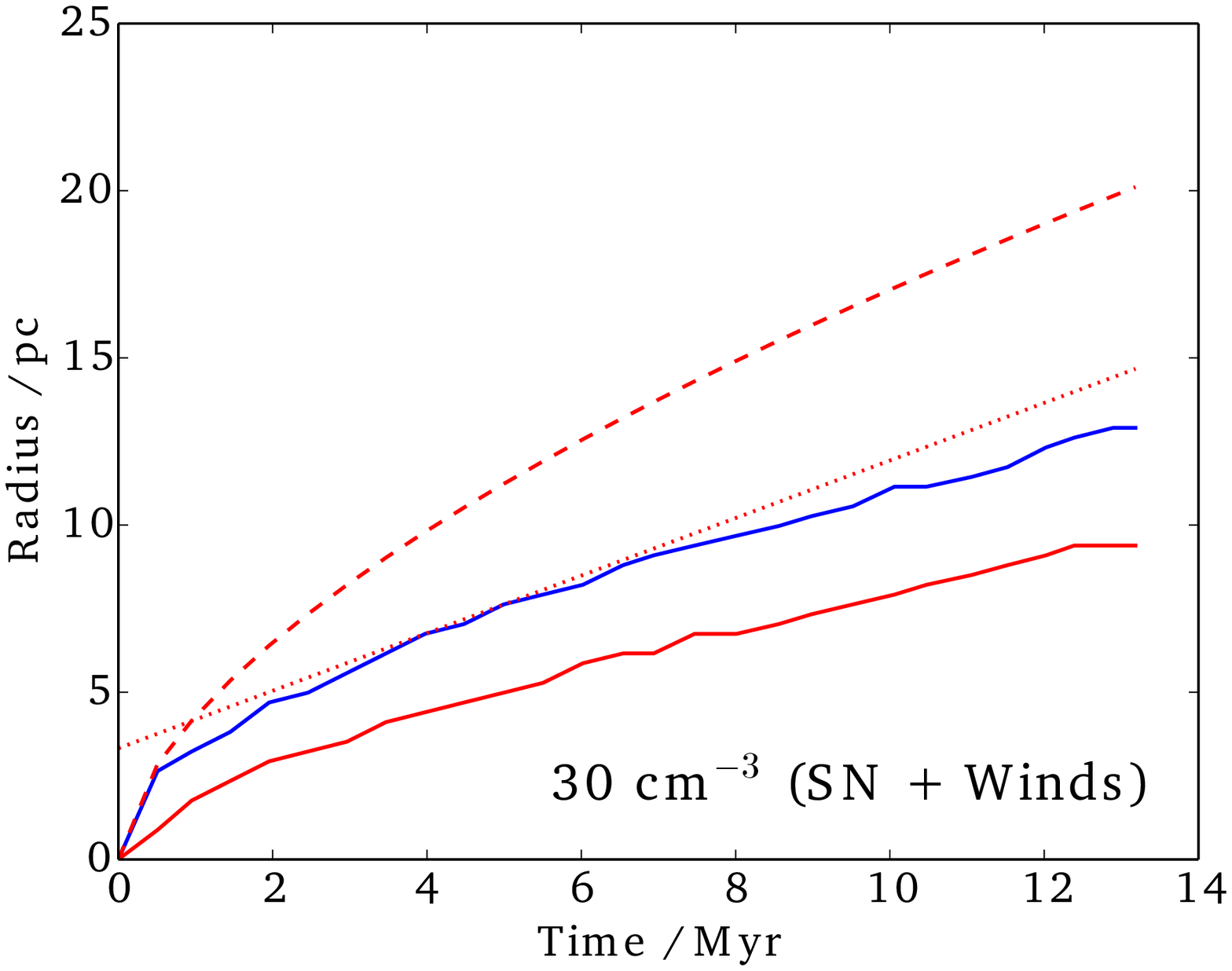}}
\centerline{\includegraphics[width=0.49\hsize]{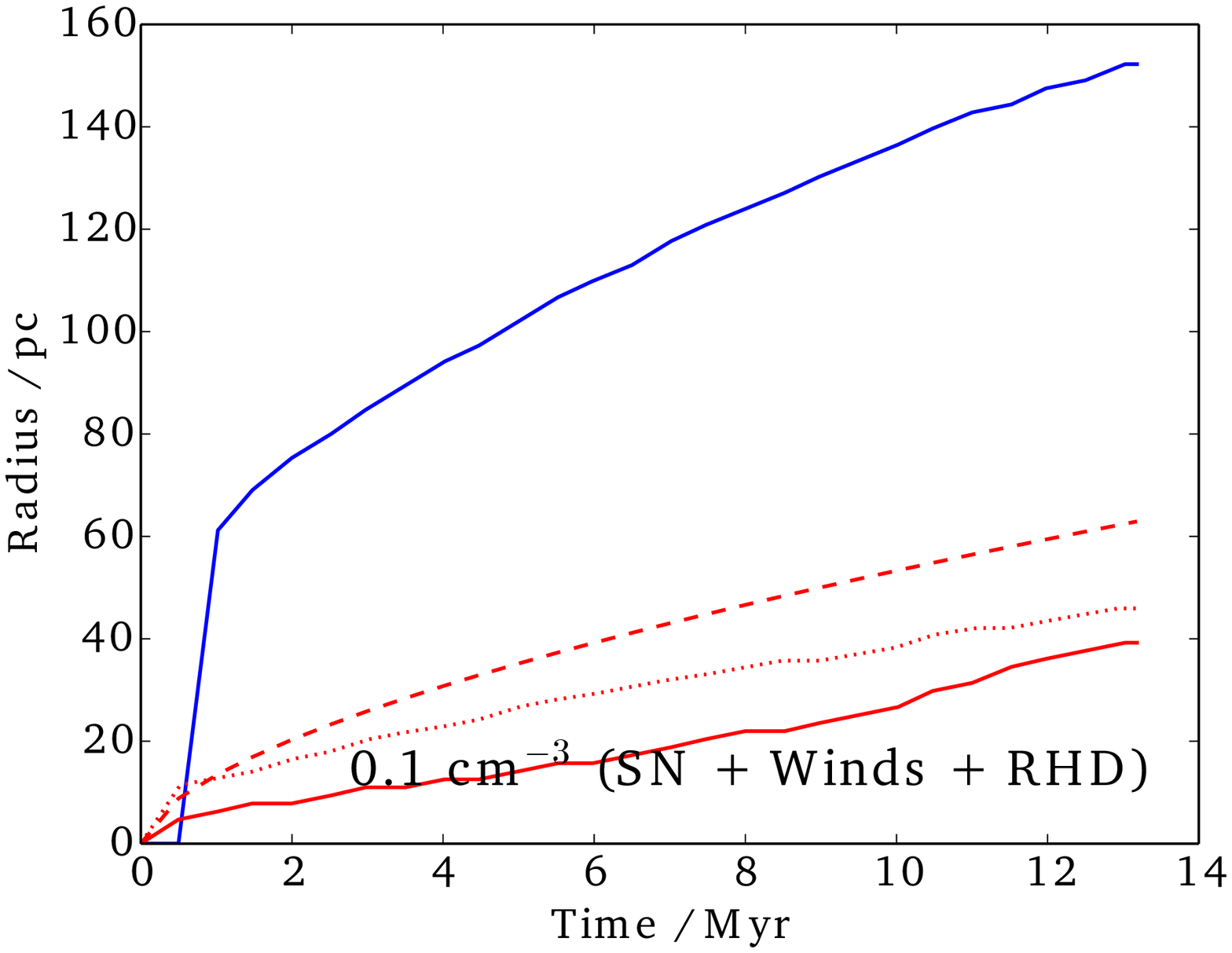} \includegraphics[width=0.49\hsize]{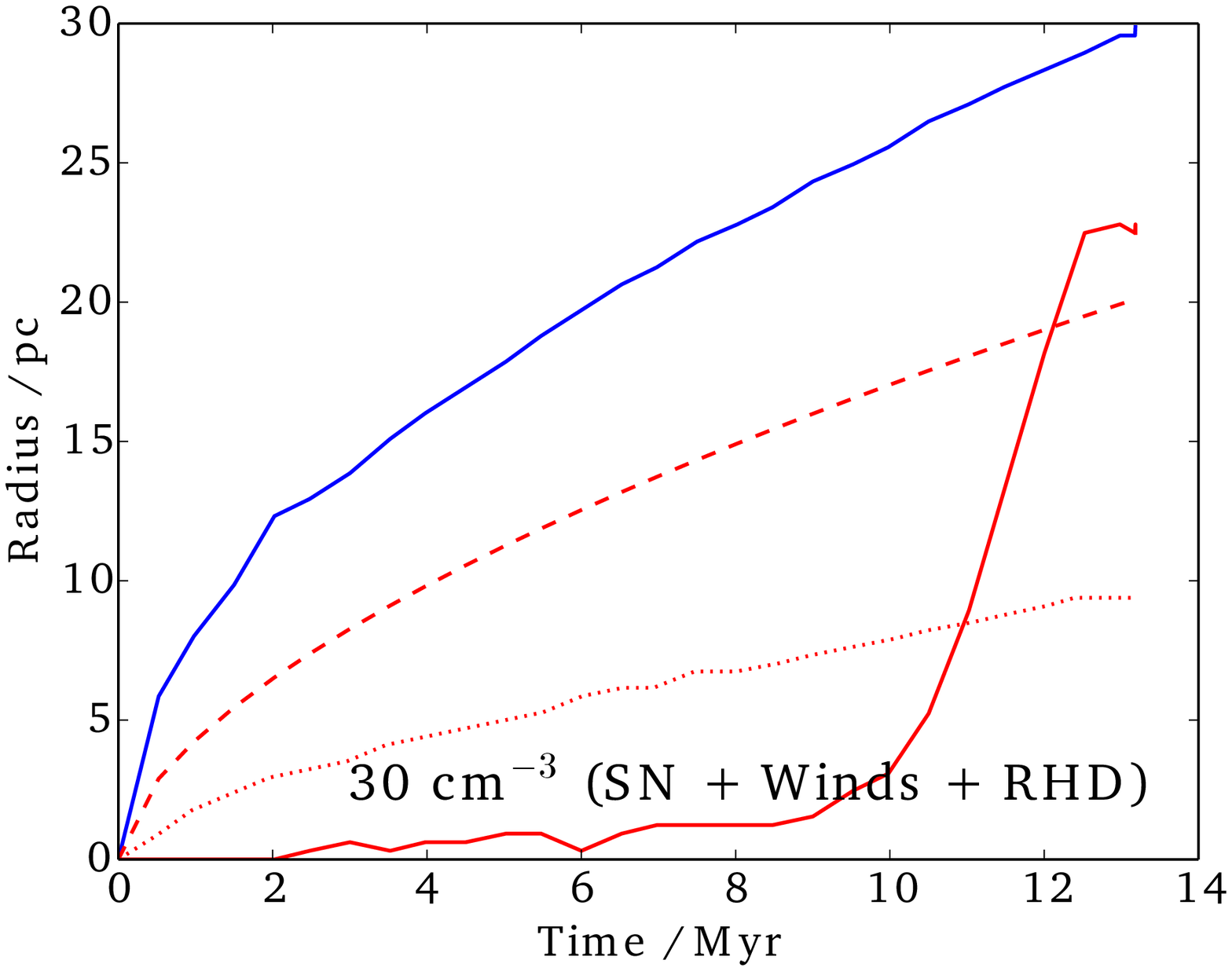}}
  \caption{Radii of wind bubbles against time in \protect\simname{N0.1ZsoSW} (top left), \protect\simname{N30ZsoSW} (top right), \protect\simname{N0.1ZsoSWR} (bottom left), \protect\simname{N30ZsoSWR} (bottom right). The radius of the shell (the maximum radius r at which $n(r) >n_0$ for a background density of $n_0$) is plotted as a solid line. In simulations including photoionisation, the shell radius lies at $r_i$, whereas in simulations without photoionisation it is found at $r_w$). The radius of the hot bubble (a heuristic value determined as the maximum radius at which T $> 2 \times 10^3$ K in runs without photoionisation, and $2 \times 10^4$ K with photoionisation) is plotted as a red line. The adiabatic solution for a wind bubble given by \protect\cite{Avedisova1972,Castor1975} is plotted as a red dashed line. In the top figures, the solution for which the internal pressure force of the bubble balances the deceleration from matter accretion by the shell is plotted as a red dotted line, positioned vertically to intersect the shell radius at 5Myr. 
  In the bottom figures, the hot bubble in the simulations containing only wind (the solid red line in the top figures) is plotted as a red dotted line.}
  \label{evolution:windradii}
\end{figure*}

The wind luminosity of our model star is significantly lower than the luminosity in ionising photons. Nonetheless, the effect of the wind is visible in the temperature and density profile, as described in \ref{evolution:overview}. In Figure \ref{methods:massloss_energy}, the wind luminosity is roughly constant until 10 Myr, when the mass loss rates increase significantly before the star explodes as a supernova at 13.1 Myr in the solar metallicity runs. In Figure \ref{evolution:windradii}, we plot the radial expansion of the stellar wind bubble for runs at 0.1 \atcc and 30 \atcc, both in the presence and in the absence of an ionisation front. In the case without photoionisation, the wind expands initially according to the adiabatic solution of \cite{Avedisova1972,Castor1975}. 

Once the structure loses a substantial portion of its energy to radiative cooling, the shell begins to decelerate. Curiously, after a few Myr in both runs \simname{N0.1ZsoSW} and \simname{N30ZsoSW}, the shells appear to reach a state where it either decelerates very slowly or not at all. From equation (54) of \cite{Weaver1977}, if the shell's acceleration is negligible we obtain a speed of the shell around the wind bubble $V_{shell}^2 = P/\rho_0$, where P is the pressure driving the shell and $\rho_0$ is the mass density in the external medium. At 5 Myr, we find that $V_{shell}$ is 2.55 km/s in \simname{N0.1ZsoSW} and 0.33 km/s in \simname{N30ZsoSW}, though in the latter case $V_{shell}$ drops below this value at later times. These profiles are overplotted on Figure \ref{evolution:windradii}. For $V_{shell}$ to be constant, the pressure at the shell also needs to be constant. In run \simname{N0.1ZsoSW}, the pressure drops throughout the main sequence of the star but is maintained at a stable value once the wind luminosity increases at late times. In run \simname{N30ZsoSW}, a similar effect occurs, although the effect is more dramatic, as the pressure inside the bubble falls by then rises by an order of magnitude once the wind luminosity increases. By the end of the lifetime of the star, the pressure inside the shell at $r_w$ in run \simname{N30ZsoSW} is far higher than the pressure at the inner edge of the wind bubble. This is because although the temperature of the shell is only around 20 K, the density of the shell is far higher than that inside the wind bubble. At this density and temperature we would have to consider cooling from molecules in order to properly determine the gas pressure. The radial expansion of the bubble is influenced to some extent by instabilities in the shell, which cause differences in the radial expansion of the shell across its surface, though from visual inspection these differences are small.

When we include the ionisation front, the wind bubble radius in \simname{N0.1ZsoSWR} expands more slowly than in the same run without photoionisation. A more dramatic effect is seen in run \simname{N30ZsoSWR}, where the wind bubble is prevented from expanding beyond 1pc until the star reaches an age of 10 Myr. This bubble is undersampled in our plots due to the small number of cells inside 1pc, leading to the bubble being identified as having zero radius for some timesteps. In addition, it is only fractionally hotter than the photoionised gas, making detection difficult. After 10 Myr, the wind bubble rapidly expands to the inner edge of the shell of the ionisation front, far beyond its extent in the simulation without photoionisation. This is because the pressure inside the ionised gas is higher than the external medium, and so the expansion of the wind is resisted, as per \cite{Weaver1977}. The pressure $P_i$ inside the ionisation front can be approximated as $2 n_i k_B T$, where $n_i$ is the number density of the ions and $T = 10^4$ K, and the factor 2 accounts for electrons (slightly higher if we include twice-ionised helium). Using equation \ref{evolution:stromgren_radius} for a constant photon flux and ionisation fraction, $P_i$ scales as $r_i^{-3/2} T$ as long as the gas remains in ionisation equilibrium. In our simulations, the pressure drops faster due to the decreasing photon flux throughout our simulation. This effect is particularly noticeable at around 10 Myr, the same time that the wind luminosity increases. As a result, the wind bubble radius grows much faster after 10 Myr.

There are a few reasons why this effect is more pronounced in the denser medium. Firstly, the final value of $r_i$ for \simname{N0.1ZsoSWR} is less than 2 \Stromgren~radii, compared to a factor of several for \simname{N30ZsoSWR} (see Figure \ref{evolution:ionradii}). As a result the pressure drops faster in the denser run from the initial value since as stated above, $P_i \propto r_i^{-3/2} T$ . Secondly, the initial pressure inside the ionised gas is much lower in \simname{N0.1ZsoSWR} than \simname{N30ZsoSWR} as the initial density is 300 times lower. This means that the wind in the diffuse case is never completely prevented from expanding. Thirdly, once the photoionised gas begins to recombine as the photon flux drops, the effect of collisional cooling in the denser gas is stronger than in the more diffuse gas. 

Hence for winds expanding inside ionisation fronts in diffuse media, which have a low initial pressure but a less marked change in pressure over time, the wind expands more slowly than in a neutral medium, but not much more. By comparison, in dense environments, the ionised gas has a high initial pressure that rapidly drops as $r_i$ expands. In this case, the wind bubble cannot expand until the front grows and the ionisation fraction drops leading to a lower temperature, at which point the wind bubble expands rapidly.


\section{The Properties of the Supernova Remnant}
\label{supernova}

\subsection{Role of Pre-Supernova Stellar Evolution}
\label{supernova:processes}

\begin{figure}
\centerline{\includegraphics[width=0.99\hsize]{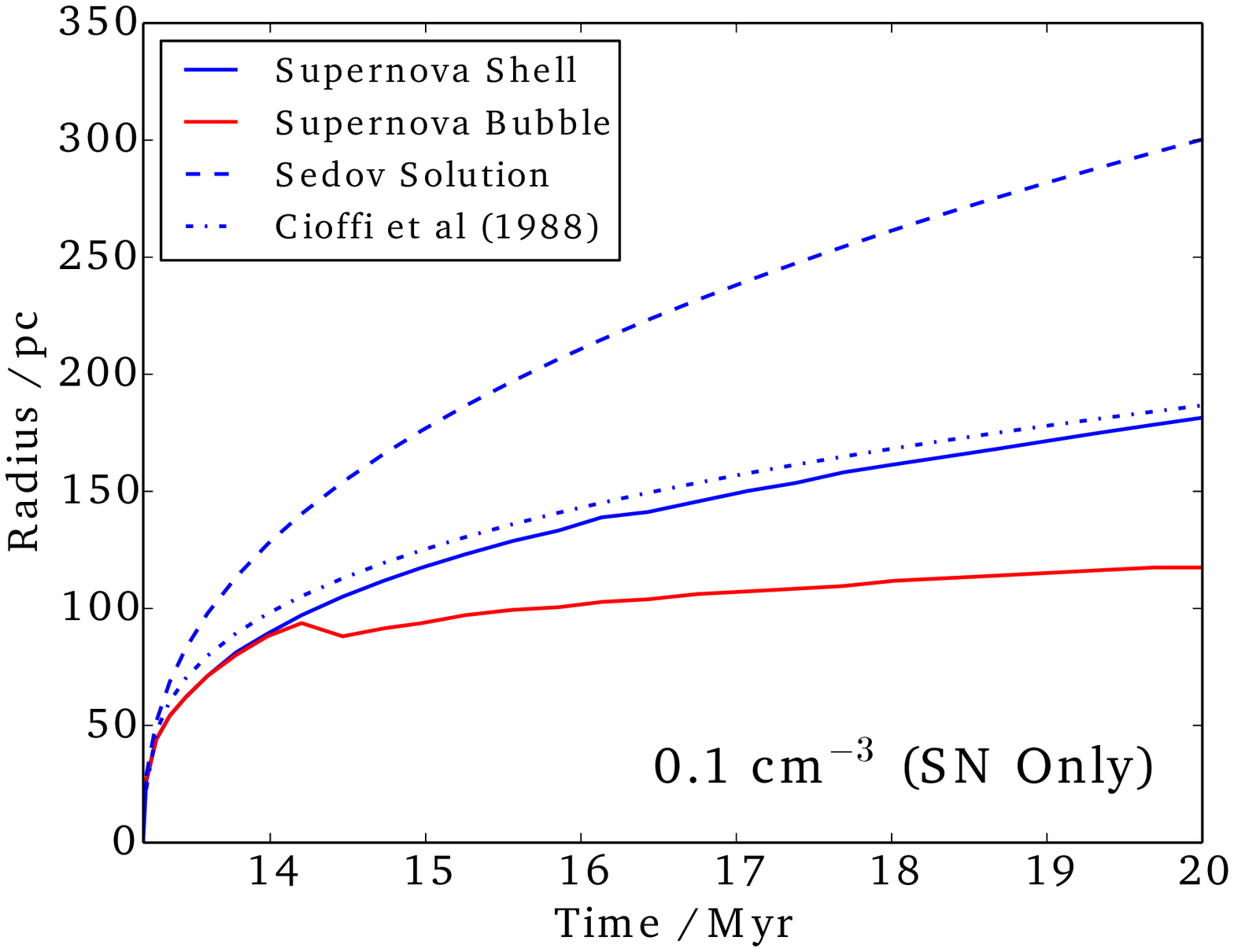}}
\centerline{\includegraphics[width=0.99\hsize]{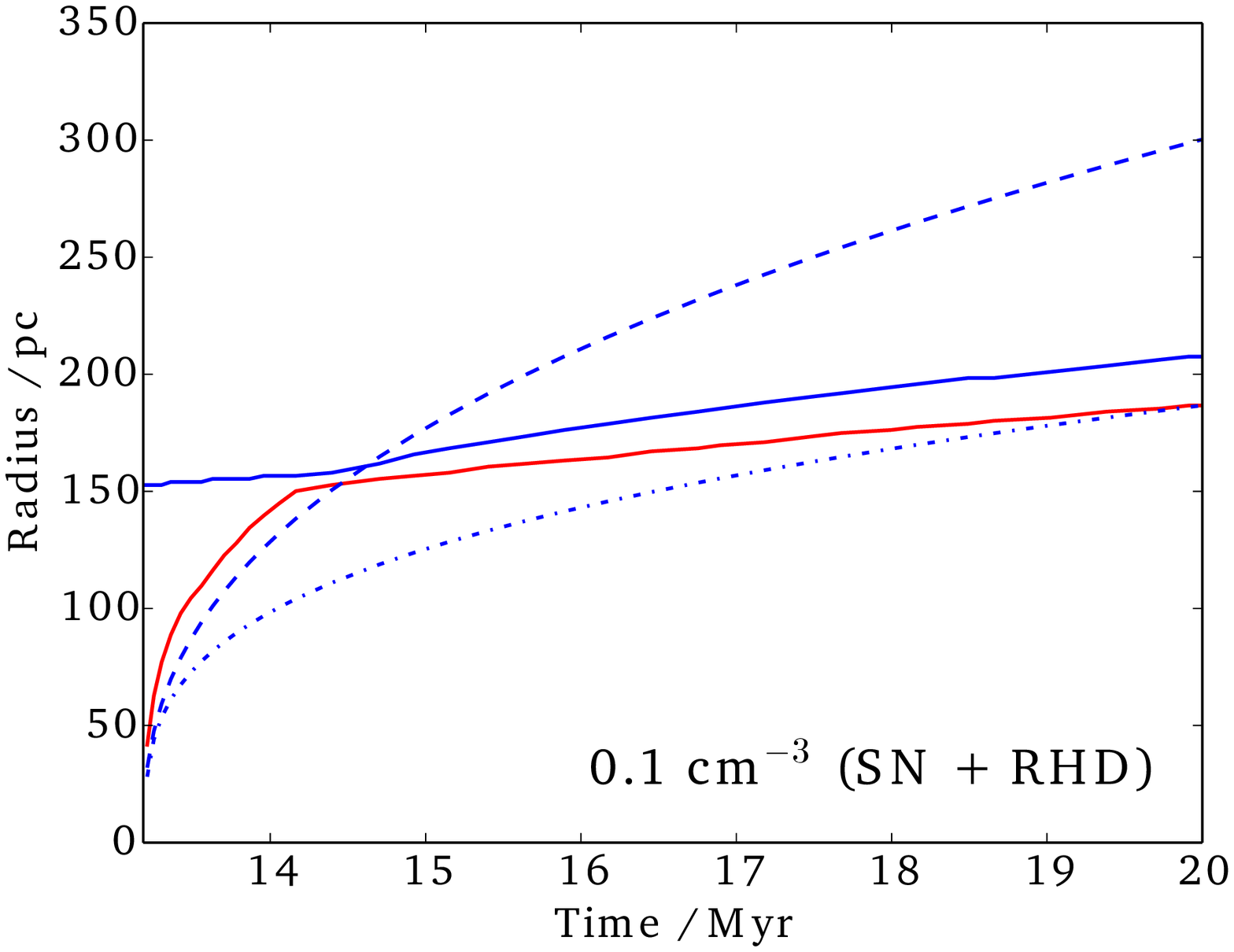}}
\centerline{\includegraphics[width=0.99\hsize]{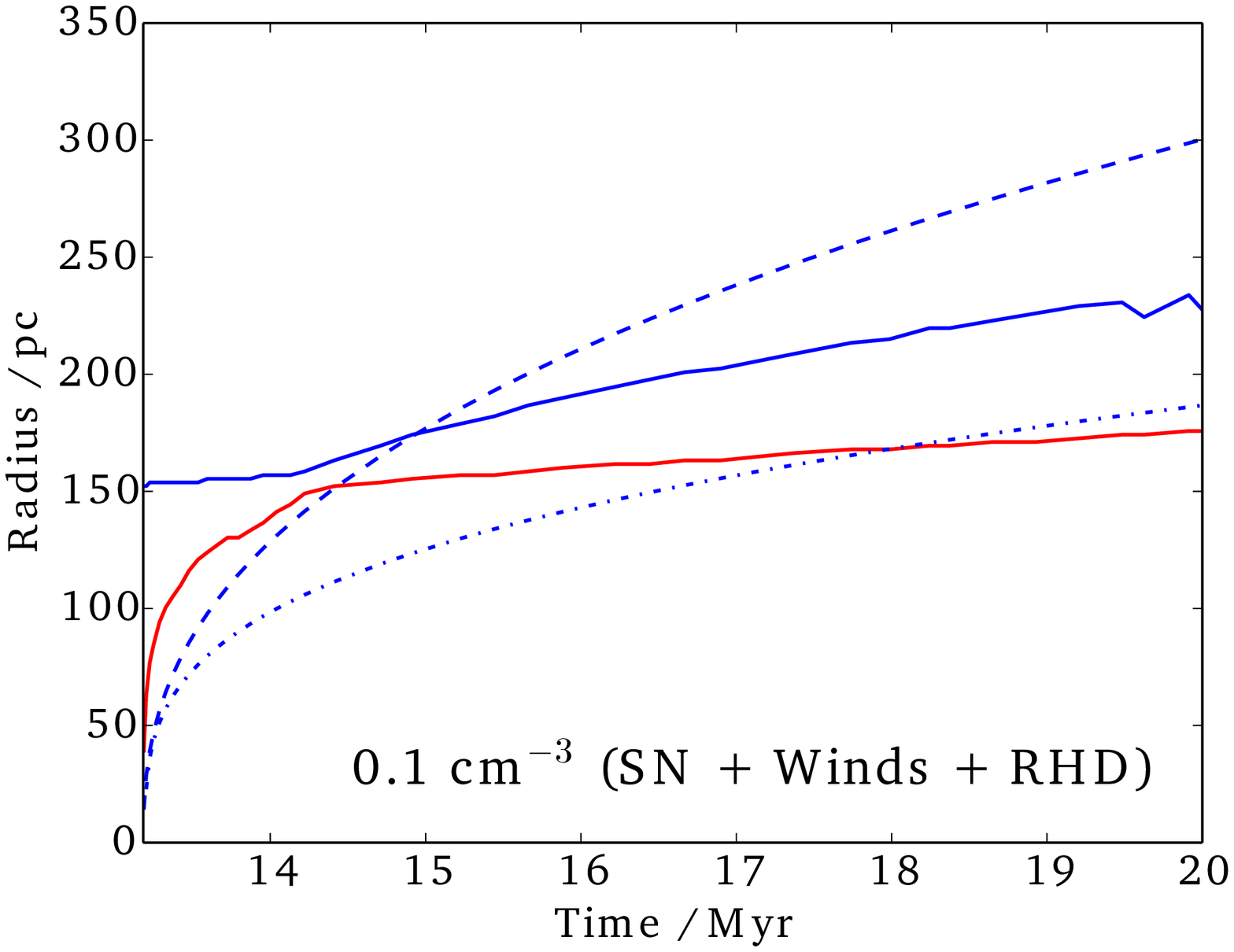}}
  \caption{The radial evolution of the supernova remnant with time. The top Figure shows \simname{N0.1ZsoS}, the middle shows \simname{N0.1ZsoSR} and the bottom shows \simname{N0.1ZsoSWR}. The radius of the shell r$_s$(t) (the maximum radius r at which n(r) $>$n$_0$ for a background density of n$_0$) is plotted as a solid blue line. The radius of the hot bubble (the maximum radius at which T $> 2 \times 10^3$ K) is plotted as a solid red line.  The Sedov solution for the given medium is plotted as a blue dashed line, while the solution found in \protect\cite{Cioffi1988} is plotted as a blue dot-dashed line.}
  \label{evolution:snradii}
\end{figure}

\begin{figure}
\centerline{\includegraphics[width=0.99\hsize]{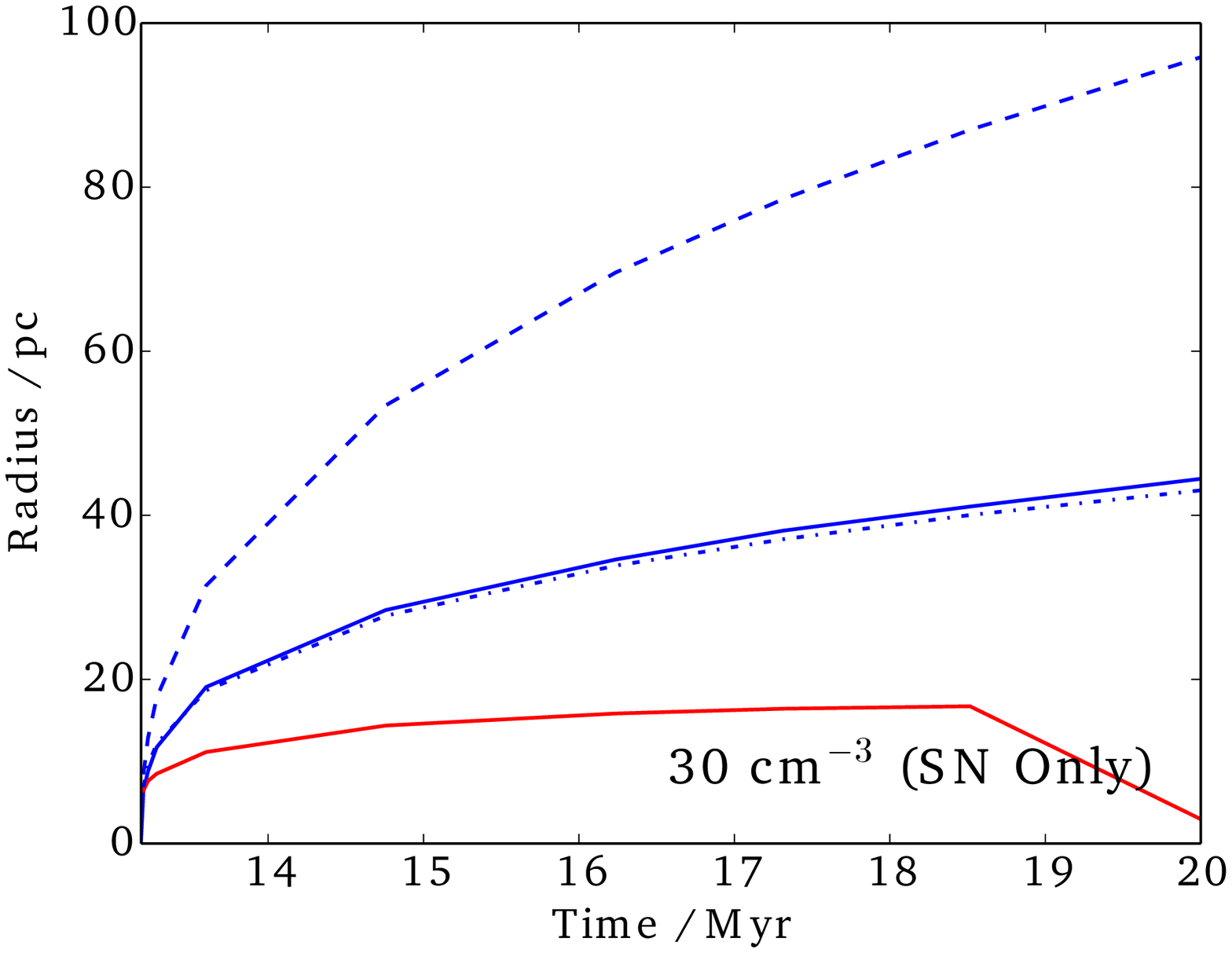}}
\centerline{\includegraphics[width=0.99\hsize]{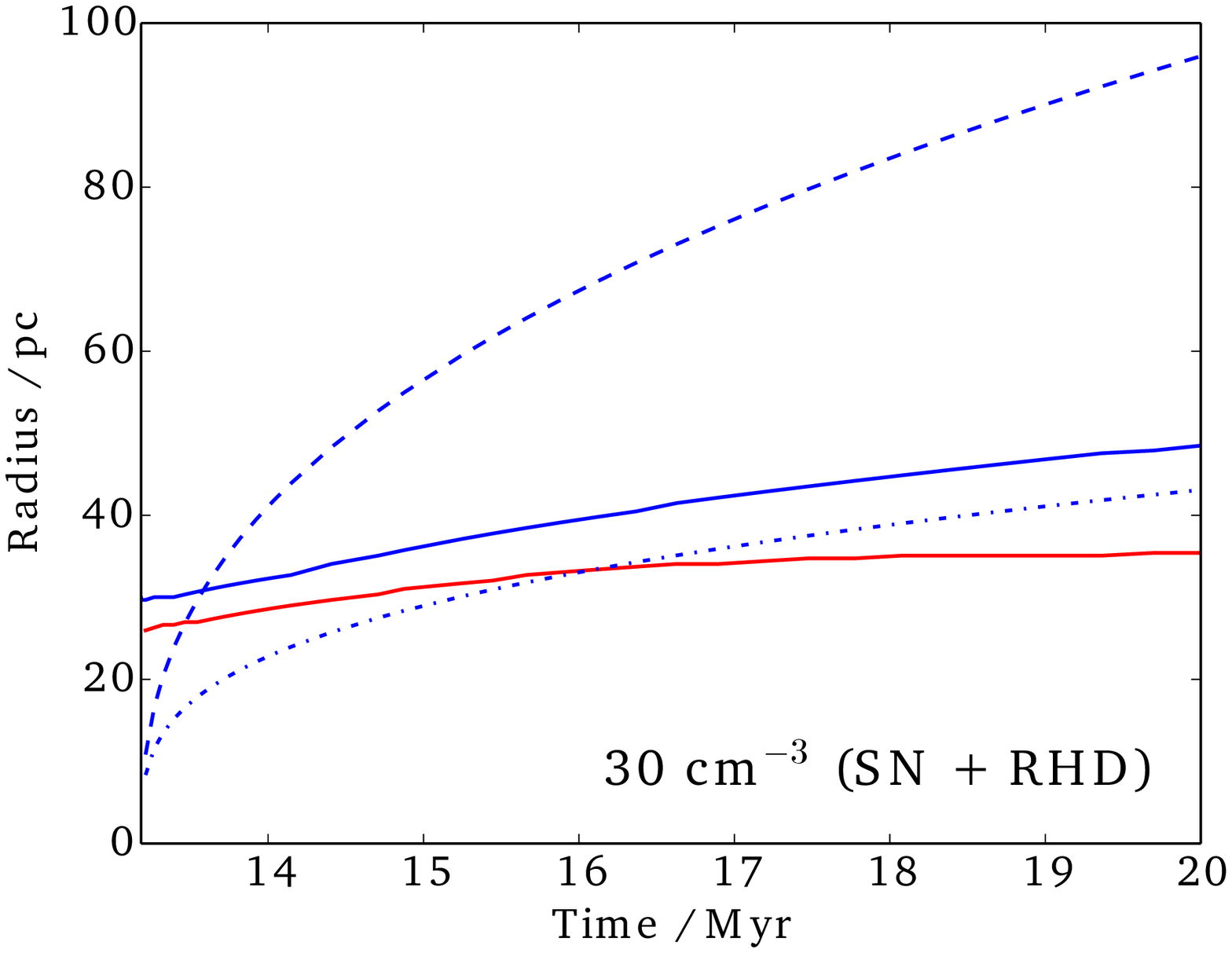}}
\centerline{\includegraphics[width=0.99\hsize]{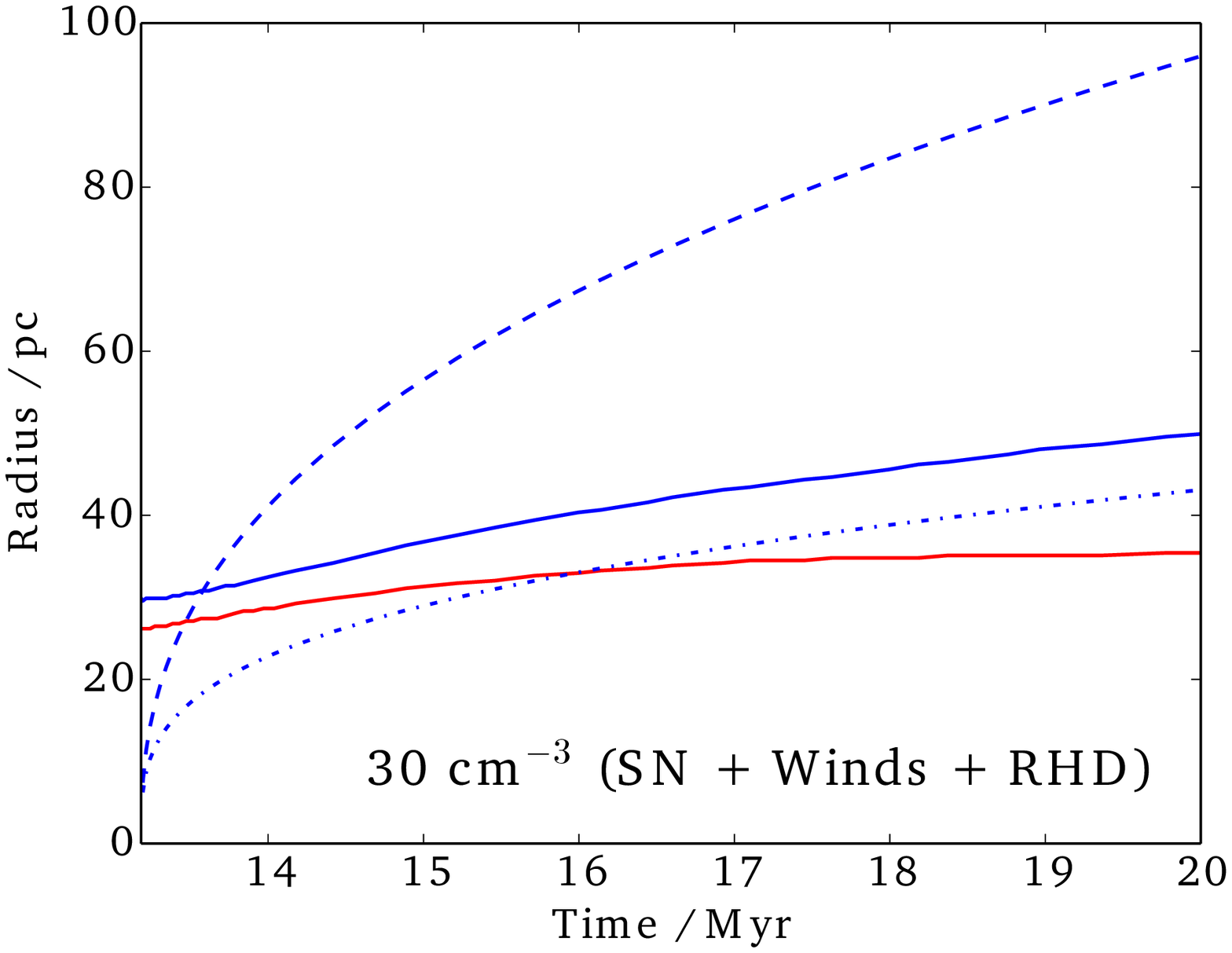}}
  \caption{As for Figure \ref{evolution:snradii}, but for \simname{N30ZsoS} (top), \simname{N30ZsoSR} (middle) and \simname{N30ZsoSWR} (bottom).}
  \label{evolution:snradii_30}
\end{figure}
%
%
%
%
%
%
%
%

At t$_{SN}$, the star explodes, creating a supernova remnant that expands into the surrounding medium. t$_{SN}$ is 13.2 Myr for the star at \Zsolar and 15.8 Myr for 0.1 \Zsolar. The radial expansion of the supernova remnant depends on the structure of the \CSM prior to the supernova. Figures \ref{evolution:snradii} and \ref{evolution:snradii_30} show the radial evolution of the supernova remnant for the runs at 0.1 and 30 \atcc at solar metallicity for simulations including just a supernova, a supernova plus photoionisation, and a supernova plus winds and photoionisation (for the sake of brevity, we omit simulations with winds but without photoionisation). The cases in which a supernova explodes into a uniform medium without stellar winds or photoionisation are well-studied in the literature. The Sedov solution \citep{Sedov1946} describes a fully-adiabatic remnant, which our results quickly deviate from as the supernova bubble loses thermal pressure to radiative cooling. A better comparison is made when we overplot an empirical formula for the radial expansion of a supernova remnant derived by \cite{Cioffi1988}, who include radiative cooling in their models. Our results lie slightly under their curve in the run \simname{N0.1ZsoS}, but the difference is not marked. In \simname{N30ZsoS} the agreement with \cite{Cioffi1988}'s formula is much better. In both cases, the shell begins to spread, i.e. the difference between the inner and outer radius of the shell grows. This is because as the bubble cools, it loses pressure, to the point where the thermal pressure in the shell is higher than that inside or outside the shell radius. In \simname{N30ZsoS}, the gas cools so rapidly that towards the end of the simulation most of the bubble falls below our threshold of $2\times10^3$ K. Once the pressure in the bubble falls below the pressure in the shell, the remnant becomes momentum-conserving, and decelerates as its kinetic energy is transferred to matter accreted by the shell from the external medium.

When we include photoionisation prior to the supernova, a large underdensity is created inside the ionisation front, and the displaced matter is piled into a shell at $r_i$. During the adiabatic phase of expansion, the shock radius follows the Sedov solution, i.e. $r_s(t) \propto n^{-1/5}t^{2/5}$. Hence for lower density environments, the supernova remnant can expand more rapidly. Typically the pressure inside the supernova bubble is much greater than that inside the HII region or wind bubble. Further, as the density is lower, the energy loss rate from radiative cooling is lower. The effect of photoionisation from the star is thus to cause the remnant to expand more rapidly and lose less energy to radiation.

The supernova blastwave reaches the ionisation front within 1 Myr in both \simname{N0.1ZsoSR} and \simname{N30ZsoSR}. At this point, the velocity of the shell drops considerably as the shock transfers its momentum to the shell. The final radius of the supernova is increased by the presence of an ionisation front. In fact, since in both cases the final radius of the ionisation front is greater than the radius of the supernova remnant in \simname{N0.1ZsoS} and \simname{N30ZsoS} after 7 Myr, the radial extent of the supernova remnant appears to be largely governed by the pre-supernova photoionisation. Adding stellar winds does not appear to significantly change the radial evolution of the remnant. The diffuse medium exhibits some radial variations from features on the surface of the shell transmitted from the wind bubble's structure by the supernova shock (see Figure \ref{evolution:slices}), but the overall radial evolution is similar. This lack of influence is due to the significantly lower energy in the wind compared to the ionising photons and supernova blast.

\begin{figure*}
\centerline{\includegraphics[width=0.48\hsize]{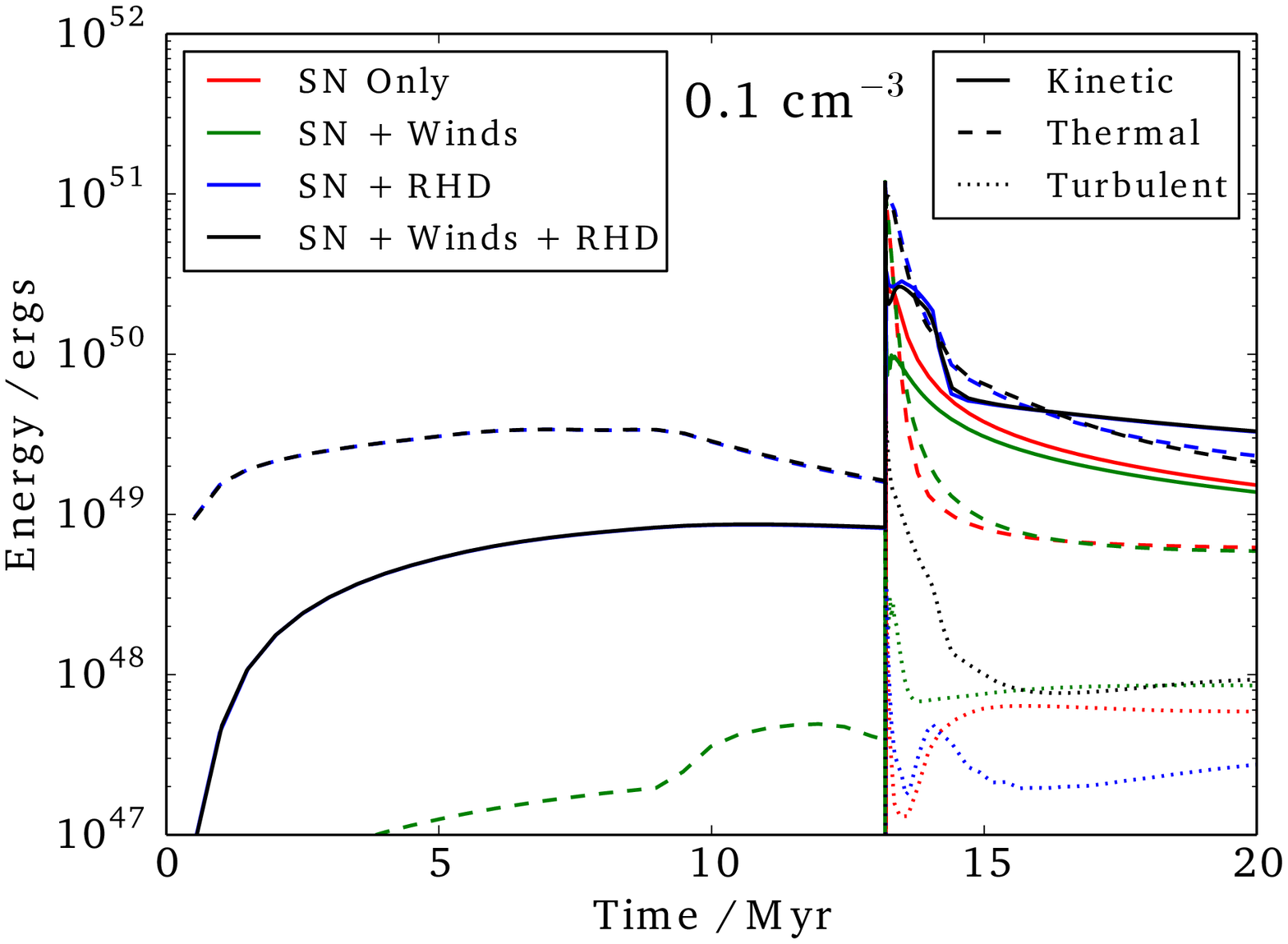}
\includegraphics[width=0.48\hsize]{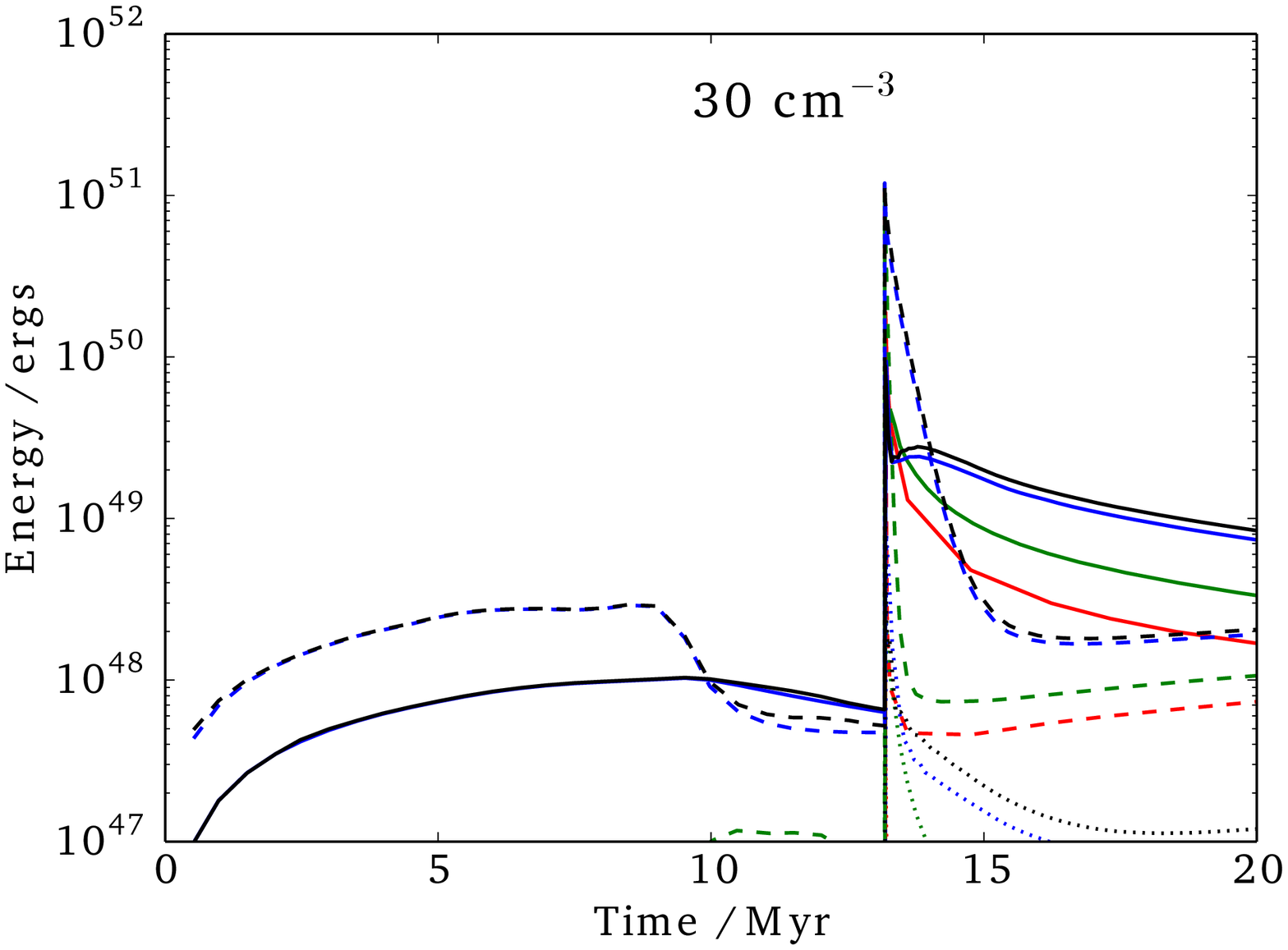}} 
\caption{Energy in the \protect\CSM over time with different physics included at solar metallicity. The left plot shows the energy in runs at 0.1\atcc. Red lines are for the run implementing only the supernova at t$_{SN}$ (labelled ``S''). Green lines are for runs including a supernova and stellar winds but no radiation (``SW''). Blue lines are for runs including a supernova and radiation but no winds (``SR''). Black lines are for runs implementing all three processes (``SWR'').  The solid lines are kinetic energy, the dashed lines are thermal energy and the dotted lines are turbulent energy, which is measured as the kinetic energy of all non-radial velocity components. The right plot is the same, but for the runs at 30\atcc.}
 \label{supernova:processes:energy}
\end{figure*}

In Figure \ref{supernova:processes:energy} we plot the evolution of the kinetic and thermal energy in each of the runs at 0.1 and 30 \atcc, adding physical processes in turn to quantify their influence on the energetics of the supernova remnant. The result of the cooling HII region due to decreasing ionising photon flux is most clearly seen in the runs at 30 \atcc. This is because the higher density leads to more efficient cooling than in the 0.1 \atcc medium. By contrast, the thermal energy from the wind bubble grows at the same time, due to a higher wind luminosity from the star, though at two orders of magnitude lower energies than that deposited by photoionisation. Once the supernova occurs, $1.2\times10^{51}$ ergs are deposited around the star as thermal energy. This quickly reaches an equipartition with kinetic energy, which sets up reverse shocks inside the $r_s$. In runs \simname{N0.1ZsoS} and \simname{N30ZsoS}, the solution quickly arrives at a thermal equilibrium, with kinetic energy dropping due to accretion of stationary matter outside $r_s$. In \simname{N30ZsoS}, the thermal energy rises after a time, an effect reported by \cite{Thornton1998}, who attribute this to accretion of thermal energy from the external medium. We find that the external medium does indeed have sufficient thermal energy to do this, although the temperature is on the order of 10 K, and so it is not clear that radiative cooling is properly captured by our cooling function at these temperatures. Adding a stellar wind to the runs at 0.1 \atcc does not change the results significantly, since the energy contribution from winds is insignificant. By comparison, winds have a significant impact on the energetics of the supernova remnant at 30 \atcc, since the early cooling rate of the supernova is reduced owing to the pre-evolved underdensity inside the wind bubble. Once radiation is added, the kinetic energy in the 0.1 \atcc runs plateaus while the shock travels through the ionised gas, then drops as the shock interacts with the shell at $r_i$. The effect of stellar winds on the kinetic energy is small in the runs at 0.1 \atcc, with a small decrease in kinetic energy due to the shock interacting with the wind bubble. By contrast, the runs at 30\atcc gain energy when winds (but not photons) are included because the denser medium makes the initial shock more succeptible to cooling than the diffuse medium.

\cite{Chevalier:1977p1817} state that only a few percent of the energy injected into the \ISM by supernovae is transmitted to the gas around it, with the rest lost to radiative cooling. We find that after 2 Myr, \simname{N0.1ZsoS} has 3\% of the initial $1.2\times10^{51}$ergs in kinetic energy (see table \ref{supernova:processes:energytable}), while \simname{N30ZsoS} only retains 0.4\%. Including photoionisation and winds has a small impact on this value at 0.1 \atcc, but \simname{N30ZsoSWR} is able to retain 1.5\% of its energy, roughly four times as much as without photoionisation, due to less efficient cooling in the low-density gas inside $r_i$.

Our \simname{N0.1ZsoS} run energy values are in good agreement with \cite{Thornton1998}, whereas we find generally lower energies for the run \simname{N30ZsoS} by a factor of a few (see table \ref{supernova:processes:energytable_tfinal} for values). Our simulations include more efficient cooling to lower temperatures, since \cite{Thornton1998} do not treat cooling below 1500 K. Similarly, \cite{Cioffi1988} do not consider cooling below $10^4$ K, whereas, as \cite{Chevalier1974} notes, much of the energy in the shell will be lost as it cools to around 10 K. Another aspect of their work is that they introduce the largest portion of their initial supernova energy as kinetic energy (as do \cite{Cioffi1988}), whereas our supernova is purely thermal (as in \cite{Chevalier1974}). It is possible that the early evolution of the shock may differ as a result, despite the fact that our simulations are adequately resolved to capture the initial cooling of the thermal blast. \cite{Cioffi1988} give a description of this early phase in the presence of a mostly-kinetic shock. \cite{Durier2012} find that the initial partition of energy in a blastwave should not affect the final result, though they do not as yet include radiative cooling in their work.

\begin{figure}
\centerline{\includegraphics[width=0.98\hsize]{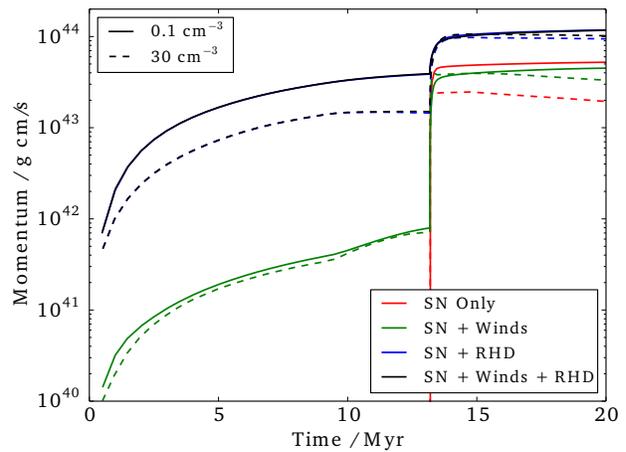}}
\caption{Total momentum over time with varying physics included. All runs are shown at \Zsolar. The solid lines show the runs at 0.1 \atcc, while dashed lines show the runs at 30 \atcc. Colours are as stated in the legend and in Figure \ref{supernova:processes:energy}.}
 \label{supernova:processes:momentum}
\end{figure}

When we look at the momentum of the gas in Figure \ref{supernova:processes:momentum}, the time evolution is somewhat simpler. It is interesting that the momentum from the stellar wind appears to be only weakly correlated with density. From \cite{Weaver1977} we can estimate the wind shell momentum as being proportional to $\rho_0^{1/5}$, assuming all the matter inside $r_w$ is displaced to $r_w$, and that the shell velocity is $\dot{r_w}$. This weak density dependence is offset by more efficient cooling in denser runs, which is not considered in \cite{Weaver1977}. Performing a similar analysis using the Spitzer solution (equation \ref{evolution:expansion_wave_spitzer}) for the ionisation front, we find that the momentum $p_i = \rho_0 r_s C_i (1 + 7/4 C_i / r_s t)^{9/7}$. Since $r_s \propto \rho_0^{-2/3}$, we find that the momentum is proportional to $1/\rho_0$ (assuming that the power of $9/7 \simeq 1$, and noting that $C_i$ is constant with density as the ionised gas is at $10^4$ K in both cases). If we assume the limit during the late evolution in which $r_s \propto n^{-1/3}$, the momentum is constant with respect to $\rho_0$. Hence the momentum in the runs at 30 \atcc is lower than the run at 0.1 \atcc, though not 300 times lower, since the runs at 0.1 \atcc fall below the Spitzer solution (see Figure \ref{evolution:ionradii}).

Once the supernova occurs and the shock reaches the edge of the HII region, there is no visible impact from the supernova shock and the shell around the HII region merging on the momentum. The final momentum in \simname{N0.1ZsoSR} is only around 10\% higher than the sum of the momentum in \simname{N0.1ZsoSR} before the supernova and the final momentum of \simname{N0.1ZsoS}, suggesting that the supernova blastwave's momentum is mostly unchanged by the HII region, and the main contribution from photoionisation is additional momentum from the shell around the photoheated gas. By contrast, the final momentum in \simname{N30ZsoSR} is much higher than the sum of the pre-supernova momentum in \simname{N30ZsoSR} and the final momentum in \simname{N30ZsoS}. This is a result of the effect discussed above, where in denser environments, the HII region lowering the density of the \CSM prior to the supernova prevents the remnant from losing a significant portion of its energy before it becomes momentum-driven. As with the results for the kinetic energy, winds have a limited impact on the momentum when photoionisation is included. The inclusion of wind but not photoionisation reduces the final momentum in \simname{N0.1ZsoSW} but raises it \simname{N30ZsoSW}. Interestingly, the momentum after the supernova in runs \simname{N0.1ZsoSW} and \simname{N30ZsoSW} converges to the same value, $4\times10^{43}$ g cm/s, suggesting that the wind has a similar effect to the HII region in determining the momentum evolution of the supernova independently of the external medium.

In the runs at 30 \atcc some 2-3 Myr after the supernova has exploded, the remnant appears to lose momentum, rather than conserving it. This is because as the shell cools, its pressure tends towards that of the external medium, causing the force resisting the expansion of the supernova remnant to become non-negligible. In the presence of an external heating term (not included in this study), we would expect this effect to be visible in the run at 0.1 \atcc as well. A final curious effect is that the post-supernova momentum is roughly constant with respect to $\rho_0$ in runs with winds or radiation but not without. We attribute this to the fact that, again, the underdensity swept out by these processes limits the early cooling of the supernova shock and hence reduces the impact of density on radiative losses. That being said, the empirical fit of \cite{Cioffi1988} finds that the final momentum should be related to density by $\rho_0^{-1/7}$, i.e. the momentum deposited in their simulations is more or less independent of density even when not in the presence of winds or photoionisation. When comparing our simulations to these authors, we find a final momentum  that is a factor of 2-3 lower than their analytical formula in both \simname{N0.1ZsoS} and \simname{N30ZsoS}, though \cite{Cioffi1988} note that even their simulations reach only 80\% of the value derived from the formula, suggesting that the analytic expression diverges by a small amount from the simulated shock. One notable difference between our work and \cite{Cioffi1988} is that the latter allows cooling in the shell only down to $10^4$ K. Another is that our shell begins to spread as the pressure inside the shell drops. Some momentum is also lost to turbulence, which we discuss below. Our values for momentum agree with the results of \cite{Walch2014} to within the spread of values found by these authors. These authors do not include stellar winds or a varying photon flux but do include a structured (non-turbulent) medium around the star. This suggests that the key effect is the evacuation of the dense gas by the HII region, and that other aspects of the pre-supernova \CSM evolution do not significantly alter the final momentum injected into the \ISM.

\begin{figure*}
\centerline{\includegraphics[width=0.48\hsize]{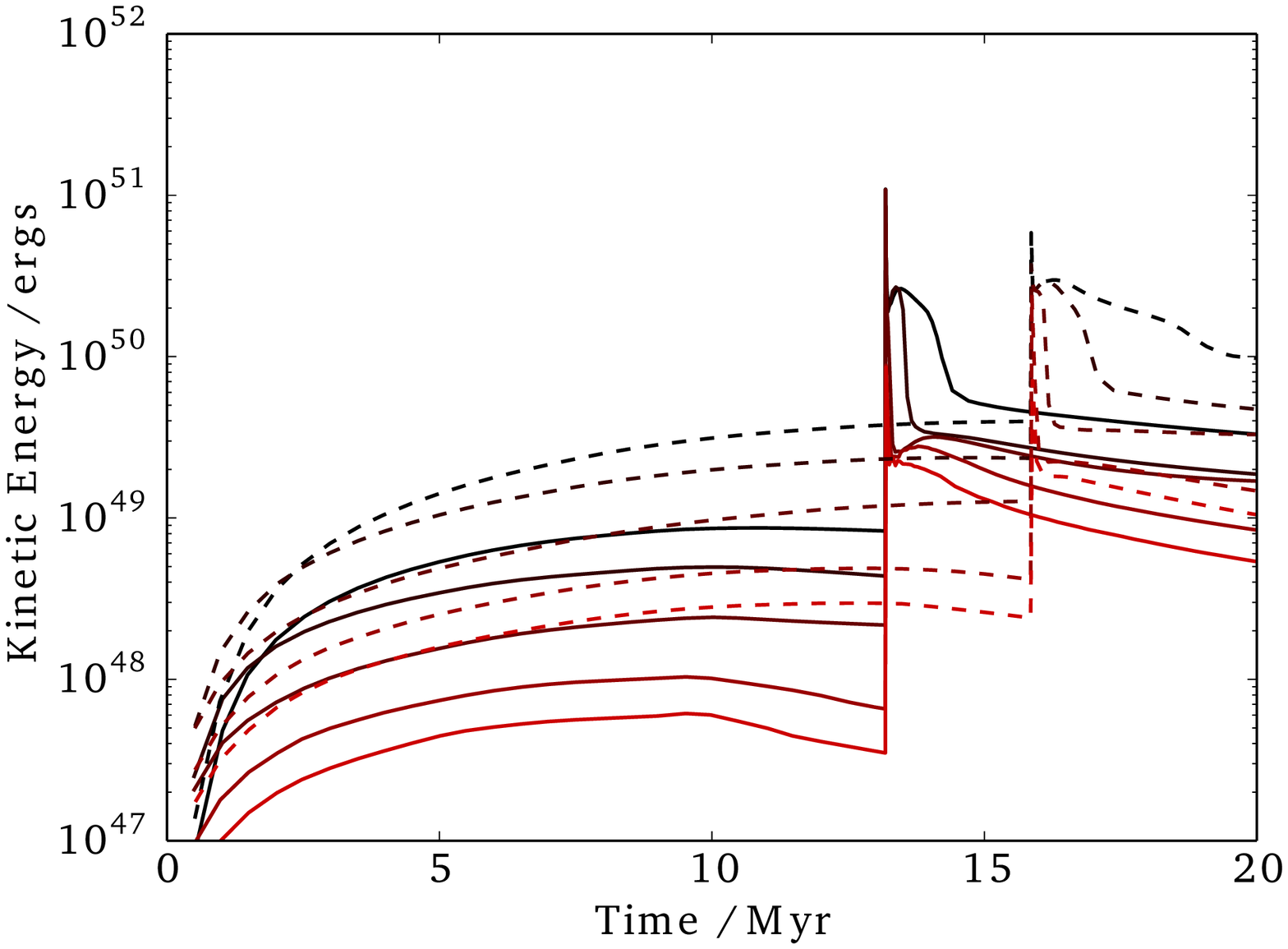}
\includegraphics[width=0.48\hsize]{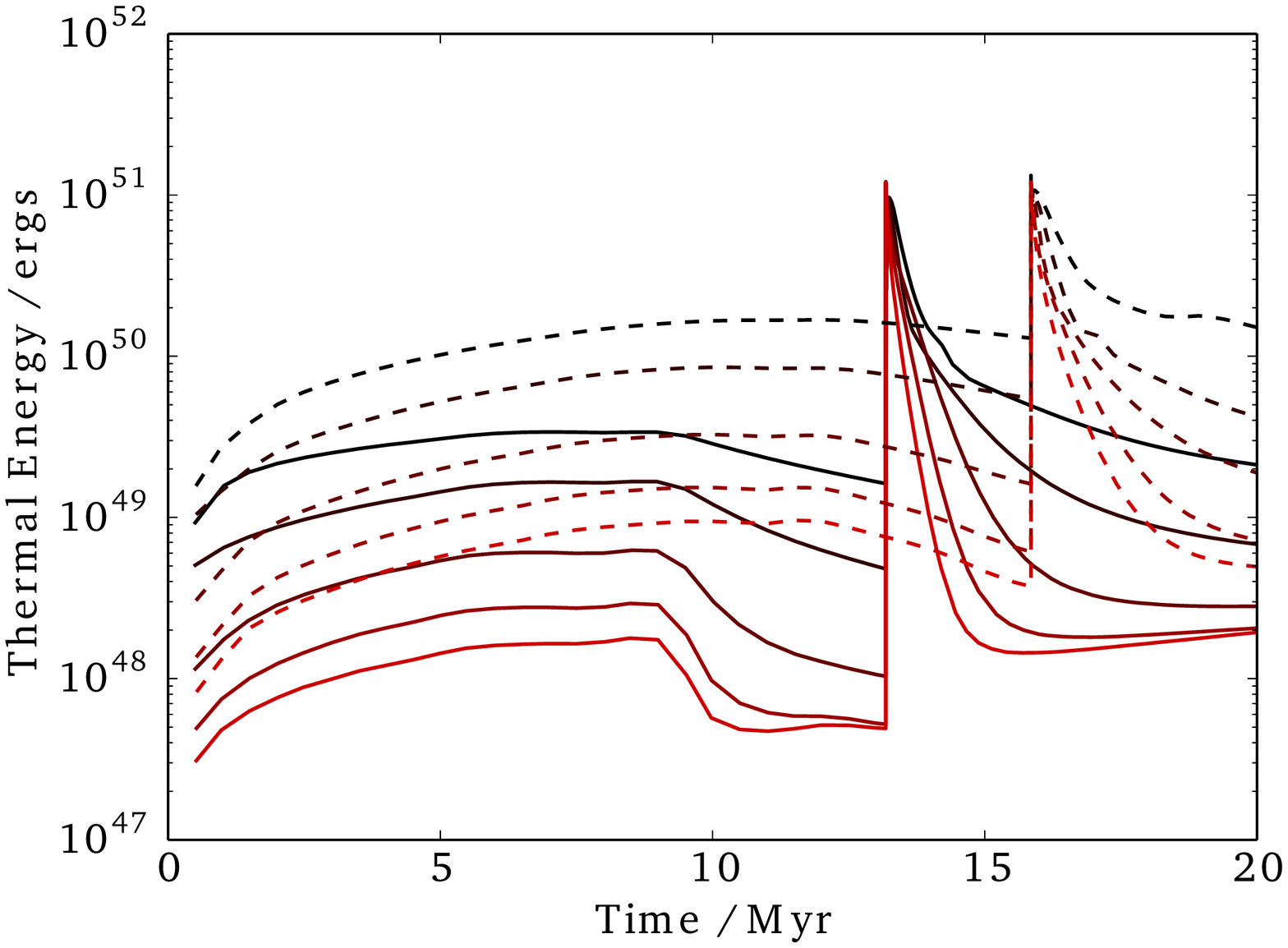}}
\centerline{\includegraphics[width=0.48\hsize]{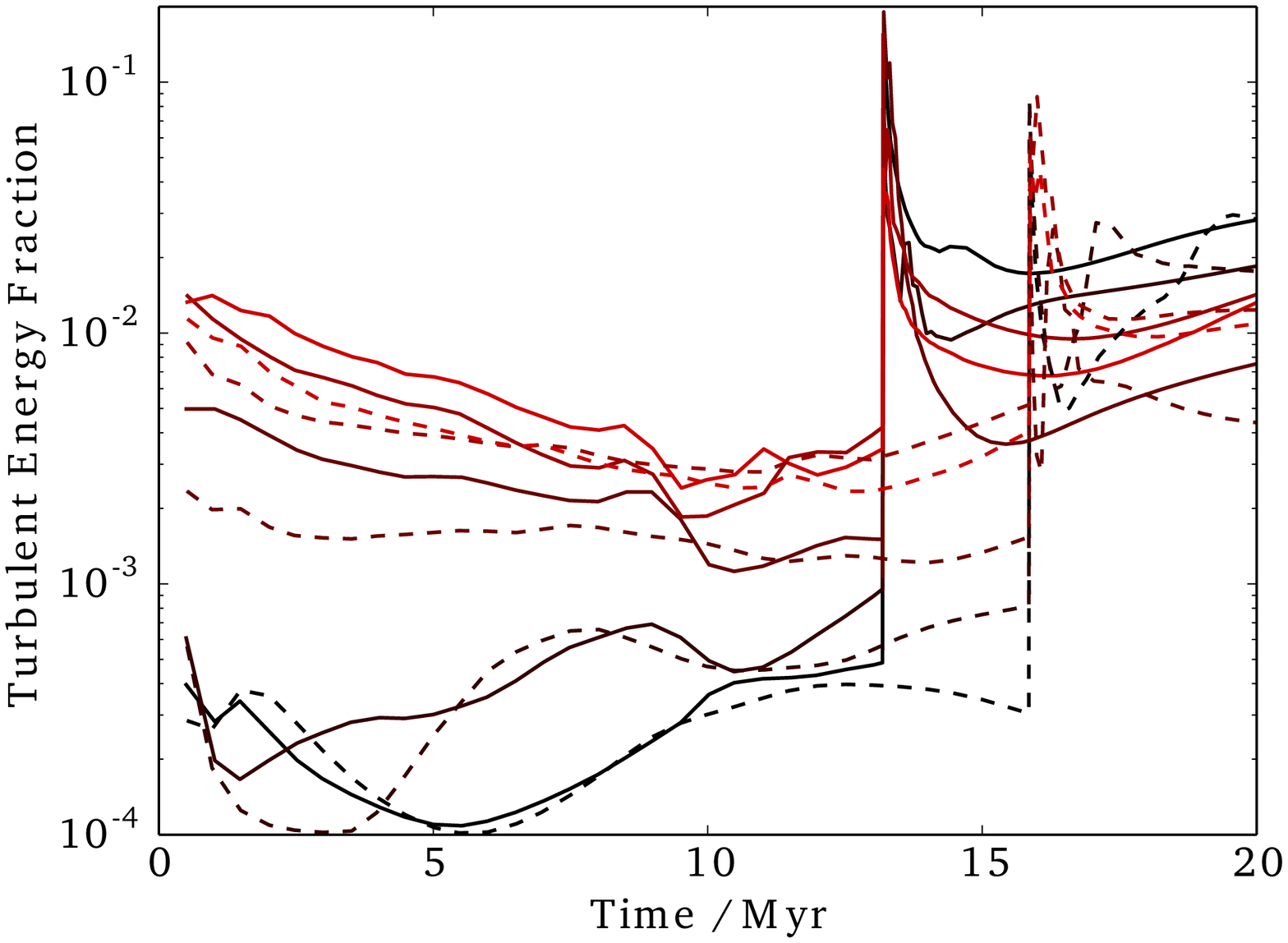}
\includegraphics[width=0.48\hsize]{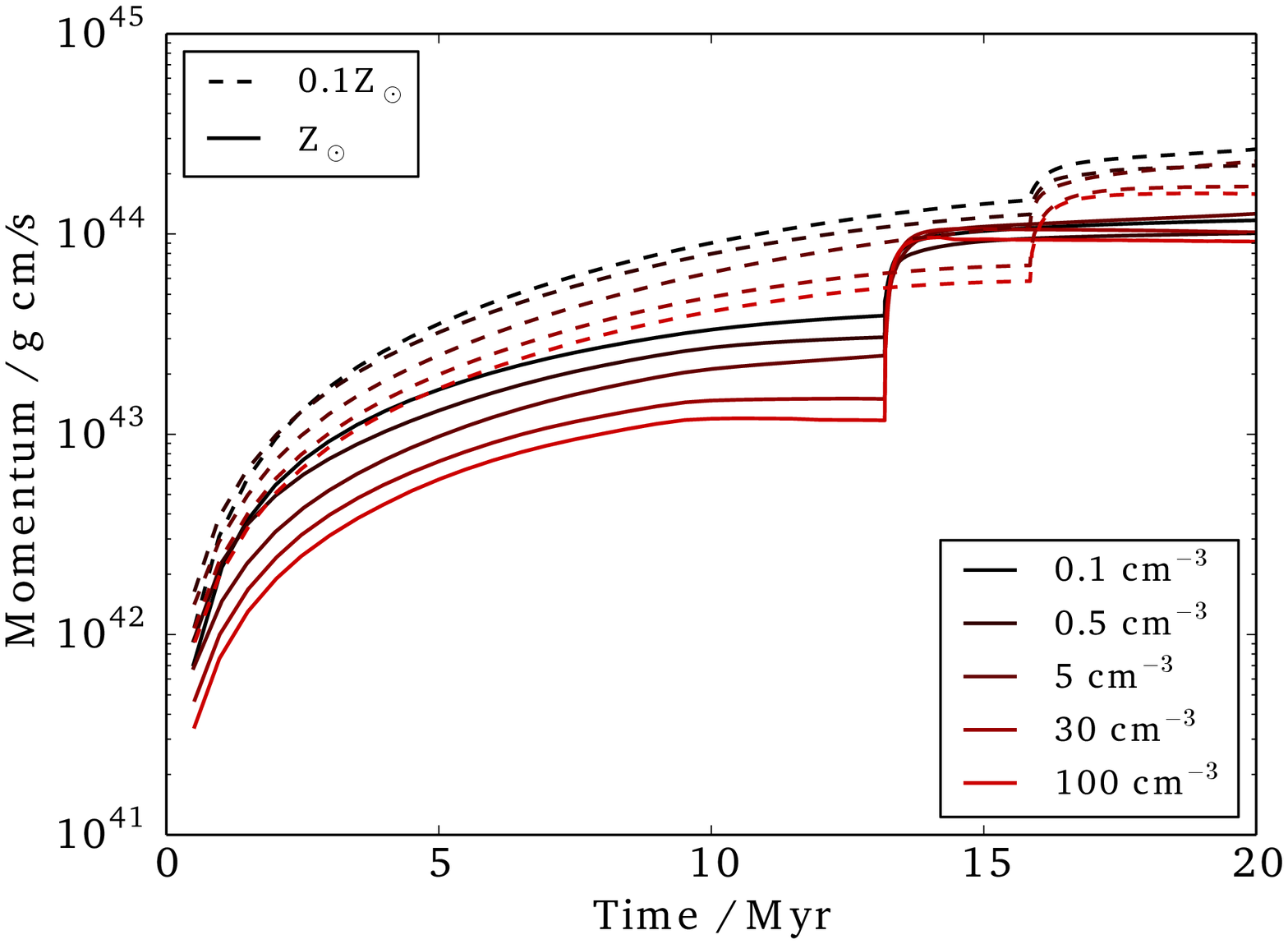}}
\caption{Evolution of the properties of the \protect\CSM over time for each of the runs containing winds and photoionisation. Shown are kinetic energy (top left), thermal energy (top right), turbulent energy fraction (bottom left) and momentum (bottom right). The \Zsolar runs are shown as a solid line and the 0.1 \Zsolar runs as dashed lines. The turbulent energy fraction is determined by measuring the fraction of the kinetic energy found in non-radial velocity components.}
 \label{supernova:environment:energetics}
\end{figure*}

In Figure \ref{supernova:environment:energetics} we plot energies and momenta for all of the runs containing both winds and photoionisation. In solar metallicity environments, these results evolve between the profiles already presented for runs \simname{N0.1ZsoSWR} and \simname{N30ZsoSWR}. At 0.1 \Zsolar, the star survives longer and has a stronger photon flux. The rate at which the supernova remnant cools is also significantly reduced. Whereas for the runs at \Zsolar we find a final momentum of $10^{44}$g cm/s mostly independent of density, at 0.1 \Zsolar we find roughly twice that value with some variation from this value as a function of density. The results suggest that the momentum from a supernova blast exploding inside a HII region is kept constant with respect to the external density by the reduced cooling inside the low-density photoheated bubble, and the small variation in the final value is dependent largely on the momentum from the shell around the ionisation front, which is larger in the low metallicity case owing to reduced cooling and increased UV flux from the star.

 In contrast to the bulk kinetic energy, turbulence is introduced mainly by the stellar winds, which transfer their aspherical structures to the supernova, which develops non-radial flows in response. We adopt a simple, robust measure for the energy in turbulence as the energy in all non-radial velocity components in the simulation. We find that turbulent flows account for less than 1\% of the kinetic energy of the system before the supernova. Similarly, once the supernova remnant has cooled, only 1\% of the kinetic energy of the system after the supernova is found in turbulent flows, in agreement with \cite{GullS.F.1973}. However, 10\% of the momentum is in non-radial velocity components. These results appear to be independent of metallicity.


\section{Conclusions}
\label{discussion}

%
%
%
%
%
%
%
%
%
%

We have described the results of a study in which a single 15\Msolar star deposits mass, momentum and energy into its surroundings. Our simulations reproduce the quasi-spherical matter distribution around a star in various environments using the M1 method for radiative transfer, with physically-motivated models for stellar winds and ionising photon fluxes and a supernova at the end of the lifetime of the star. As the star evolves, the flux of ionising photons decreases as the stellar radius grows and the surface temperature drops. As a result, we find that for the more diffuse media the ionisation front deviates from the \cite{SpitzerLyman1978} solution, and suggest an alternative model that takes into account the recombination time and decreasing ionisation fraction inside the ionisation front. As in previous work, the amount of energy in ionising photons transferred to the \ISM is on the order of 0.1-0.01\%, with most of it lost to radiative cooling from recombination processes.

The expansion of wind bubbles is highly sensitive to changes in the wind luminosity and ionising photon flux throughout the stellar lifetime. Their evolution is determined by the pressure balance between the edge of the wind bubble and the inside of the HII region. This balance changes throughout the simulation as the HII region expands. While the wind luminosity grows in the final stages of stellar evolution, the photon flux drops and the HII bubble leaves ionisation equilibrium. In the denser environments, this balance means that the wind bubble is even effectively prevented from forming until the \HGB phase, at which point it expands rapidly out to the edge of the HII bubble. The photons from the star then heat the unshocked wind inside to $10^4$ K, leading to a structure that appears to be a wind bubble with a smaller HII region embedded within it. If this effect occurs in more general cases, it could be an important consideration when modelling the temperature and structure of HII regions. Even in uniform environments, the structure of the \CSM prior to the supernova is dependent on the interplay between the varying winds and UV flux and the initial gas density. \rev{We find that the winds from stars of the mass studied in this paper do not significantly contribute to the energy of the \ISM, though we have not considered winds from more massive stars. \cite{Agertz2013} tabulate the energy from winds in a population of stars and find a much higher value, suggesting that Wolf-Rayet winds from more massive stars than the one modelled here may be more significant.}

The supernova explodes inside an underdensity surrounded by a dense shell carved out by ionising photons and winds from the supernova progenitor star. The ionisation front both provides momentum to the \ISM and reduces the loss of energy in the supernova shock from radiative cooling due to this underdensity. The former process is more important in diffuse media and the latter is more important in denser media. For solar metallicity environments, a final value of $10^{44}$ g cm/s is found for the momentum of the remnant, and $2 \times 10^{44}$ g cm/s for 10\% solar metallicity environments, with more variation in the lower metallicity runs with different initial densities. From our results, it appears that the supernova blast adds more or less the same amount of momentum to the \ISM independent of density if it occurs within a photoionised bubble, while some variance in the final momentum is caused by the momentum in the shell around the ionisation front prior to the supernova. Our results (without photoionisation or winds) agree well with the radial expansion of the supernova remnant found in \cite{Cioffi1988} and the energies in \cite{Thornton1998}, but our momentum values are somewhat lower than the expressions given by \cite{Cioffi1988}. We posit that this is due to simplifications made by their analytic function and more efficient cooling in our simulations. By contrast, we report good agreement with \cite{Walch2014}, who do include photoionisation, despite differences in our simulation setups, suggesting that for a single star $10^{44}$ g cm/s is a good estimate of the momentum addes to the \ISM a solar-metallicity star. In appendix \ref{tables} we provide lists of numerical values from our simulation. This seems to suggest that while the structure of the remnant is sensitive to the physical model used and the initial conditions, the final momentum added to the \ISM is more robust to changes in the simulation setup. However, neither of these works includes a turbulent \ISM, which could become important in modelling the propagation of shocks from stars.

\rev{Turbulence in the remnant, approximated as the energy in non-radial flows, is calculated to be around 1\% of the kinetic energy and 10\% of the momentum, depending on the density and metallicity of the gas around the star. We thus do not expect a great deal of divergence between our 3D work and a 1D spherically symmetric simulation with the same initial conditions. However, without doing the 3D experiment we cannot be sure that 1D spherically symmetric simulations would be sufficient for modelling the explosions of stars of different stellar masses, which could seed larger instabilities and therefore give rise to more turbulence. For more realistic environments containing turbulent, multiphase fluid, self-consistent star formation and a galactic disk structure, the spherically symmetric approximation breaks down and 3D simulations become unavoidable. We discuss some implications of this below.}

There are a number of limitations to this work that should be considered. For one thing, we only simulate one star (albeit at two metallicities), rather than multiple stellar masses across the full \IMF from 8 \Msolar upwards. The wind luminosities and spectra of massive stars vary greatly depending on their initial mass, metallicity and rotation period, as well as multiplicity for the case of interacting binaries. Supernova energies for the most massive stars are either much larger or much smaller than the fiducial $10^{51}$ergs \citep{Nomoto2003}. Another consideration is whether supernovae add more momentum and energy to the \ISM when they explode as part of a superwind, in which a succession of supernovae drive the expansion of a superbubble. This suggestion has recently been explored in models by \citep{Sharma2014,Keller2014}, who find substantial differences compared to results using isolated supernovae.

The environment around the star is also more complex than the uniform medium modelled in our work. It is not clear whether supernovae are more likely to explode in denser environments, in which stars are formed and which can live longer than the massive stars that form in them \citep{Hennebelle2012}, or more diffuse environments that make up most of the \ISM by volume. In addition to being multiphase, the \ISM is turbulent, which adds an effective pressure to the medium that resists propagating stellar shocks \citep{Raga2012}. Recent simulations by \cite{Tremblin2014} seek to address this by simulating ionisation fronts in both 1D and 3D in the presence of turbulence. As stated in the introduction, there may be a case for including radiation pressure in future work. Stellar motions with respect to the \ISM, not covered in this work, can also lead to features such as bow shocks. Magnetic fields and thermal conduction can also play a role in the \ISM, although in our simple near-spherical setup magnetic fields are not expected to have a great effect (see, for example, \cite{Chevalier1974}), while conduction at the Field criterion requires a much higher resolution than that available in our runs. Various of these outstanding issues will be addressed in future works.

\section{Acknowlegements}
\label{acknowledgements}

 We would like to thank Philipp Podsiadlowski, Yohan Dubois and Patrick Hennebelle for useful comments and discussions during the writing of this paper, as well as the anonymous referee for suggested improvements to the text. We warmly thank Jonathan Patterson for his help in producing the results in this paper. The simulations presented here were run primarily on the DiRAC facility jointly funded by STFC, the Large Facilities Capital Fund of BIS and the University of Oxford. Additional simulations were run on the CC-IN2P3 Computing Center (Lyon/Villeurbanne - France), a partnership between CNRS/IN2P3 and CEA/DSM/Irfu.SG has received funding from a Cosmocomp fellowship and the European Research Council under the European Community's Seventh Framework Programme (FP7/2007-2013 Grant Agreement no. 306483 and no. 291294). JR is funded by the European Research Council under the European Union’s Seventh Framework Programme (FP7/2007-2013) / ERC Grant agreement 278594-GasAroundGalaxies, and the Marie Curie Training Network CosmoComp (PITN-GA-2009-238356). JB acknowledges support from the ANR BINGO project (ANR-08-BLAN-0316-01). JD and AS's research is supported by funding from Adrian Beecroft, the Oxford Martin School and the STFC. 

 \bibliographystyle{mn2e}
 \bibliography{supernovapaper}

 \appendix
 \section{Tabulated Energies and Momenta}
 \label{tables}

\begin{table*}
\begin{tabular}{l c c c c c c c c c}
\textbf{Runs} & \textbf{R$_{tot}$} & \textbf{R$_{kin}$} & \textbf{R$_{th}$} & \textbf{R$_{turb}$} & \textbf{S$_{kin}$} & \textbf{B$_{th}$} & \textbf{S$_{mom,bulk}$} & \textbf{S$_{mom,turb}$} \\ 
\hline
\simname{N0.1ZsoS} & 49.624 & 49.536 & 48.889 & 47.800 & 49.519 & 48.133 & 43.673 & 42.759 \\ 
\simname{N0.1ZsoSW} & 49.572 & 49.457 & 48.938 & 47.888 & 49.409 & 48.560 & 43.564 & 42.769 \\ 
\simname{N0.1ZsoSR} & 50.027 & 49.685 & 49.764 & 47.327 & 49.678 & 49.704 & 44.024 & 42.675 \\ 
\simname{N0.1ZsoSWR} & 50.037 & 49.687 & 49.780 & 47.955 & 49.678 & 49.715 & 44.013 & 43.040 \\ 
\hline
\simname{N30ZsoS} & 48.607 & 48.556 & 47.654 & 46.223 & 48.508 & 46.165 & 43.339 & 42.188 \\ 
\simname{N30ZsoSW} & 48.946 & 48.907 & 47.876 & 46.609 & 48.849 & 46.637 & 43.562 & 42.412 \\ 
\simname{N30ZsoSR} & 49.264 & 49.209 & 48.344 & 47.137 & 49.188 & 47.855 & 43.979 & 42.887 \\ 
\simname{N30ZsoSWR} & 49.324 & 49.269 & 48.398 & 47.299 & 49.249 & 47.962 & 44.015 & 42.974 \\
\hline
\end{tabular}
  \caption{Table of energies and momenta calculated from each simulation in the runs at 0.1 and 30 \atcc 2 Myr after the supernova. As in \protect\cite{Thornton1998}, ``R'' refers to the remnant, i.e. the whole structure around the star, ``S'' refers to the shell, and ``B'' refers to the hot bubble. The subscripts ``tot'', ``kin'', ``th'' and ``turb'' refer respectively to the total, kinetic, thermal and turbulent energy. All energy values are in log$_{10}$(ergs). The subscripts ``mom,bulk'' and ``mom,turb'' refer to the bulk momentum and the momentum in turbulent flows respectively. All momentum values are in log$_{10}$(g cm/s).}
\label{supernova:processes:energytable}
\end{table*}

\begin{table*}
\begin{tabular}{l c c c c c c c c c}
\textbf{Runs} & \textbf{t$_{f}$/Myr} & \textbf{R$_{tot}$} & \textbf{R$_{kin}$} & \textbf{R$_{th}$} & \textbf{R$_{turb}$} & \textbf{S$_{kin}$} & \textbf{B$_{th}$} & \textbf{S$_{mom,bulk}$} & \textbf{S$_{mom,turb}$} \\ 
\hline
\simname{N0.1ZsoS} & 0.669 & 50.042 & 49.964 & 49.257 & 47.295 & 49.952 & 48.835 & 43.662 & 42.216 \\ 
\simname{N0.1ZsoSW} & 0.144 & 50.590 & 49.973 & 50.470 & 48.439 & 49.938 & 50.451 & 43.445 & 42.461 \\ 
\simname{N0.1ZsoSR} & 2.720 & 49.958 & 49.653 & 49.661 & 47.293 & 49.645 & 49.591 & 44.035 & 42.701 \\ 
\simname{N0.1ZsoSWR} & 2.176 & 50.015 & 49.676 & 49.750 & 47.927 & 49.665 & 49.682 & 44.016 & 43.037 \\ 
\hline
\simname{N30ZsoS} & 0.101 & 49.607 & 49.598 & 47.951 & 46.591 & 49.589 & 46.421 & 43.370 & 41.795 \\ 
\simname{N30ZsoSW} & 0.108 & 49.911 & 49.659 & 49.556 & 47.991 & 49.597 & 49.536 & 43.558 & 42.666 \\ 
\simname{N30ZsoSR} & 0.190 & 50.442 & 49.353 & 50.405 & 48.012 & 49.253 & 50.398 & 43.822 & 42.777 \\ 
\simname{N30ZsoSWR} & 0.073 & 50.834 & 49.551 & 50.811 & 48.200 & 48.985 & 50.808 & 43.567 & 42.603 \\ 
\hline
\end{tabular}
  \caption{As for table \ref{supernova:processes:energytable} but sampled at t$_f$, which is defined as 13 times the time after the supernova at which the total luminosity from radiative cooling is at a maximum (see \protect\cite{Thornton1998}).}
\label{supernova:processes:energytable_tfinal}
\end{table*}


\begin{table*}
\begin{tabular}{l c c c c c c c c c}
\textbf{Runs} & \textbf{R$_{tot}$} & \textbf{R$_{kin}$} & \textbf{R$_{th}$} & \textbf{R$_{turb}$} & \textbf{S$_{kin}$} & \textbf{B$_{th}$} & \textbf{S$_{mom,bulk}$} & \textbf{S$_{mom,turb}$} \\ 
\hline
\simname{N0.1ZsoSWR} & 50.037 & 49.687 & 49.780 & 47.955 & 49.678 & 49.715 & 44.013 & 43.040 \\ 
\simname{N0.5ZsoSWR} & 49.771 & 49.470 & 49.470 & 47.525 & 49.454 & 49.401 & 43.959 & 42.869 \\ 
\simname{N5ZsoSWR} & 49.571 & 49.435 & 49.000 & 47.000 & 49.416 & 48.883 & 44.031 & 42.725 \\ 
\simname{N30ZsoSWR} & 49.324 & 49.269 & 48.398 & 47.299 & 49.249 & 47.962 & 44.015 & 42.974 \\ 
\simname{N100ZsoSWR} & 49.152 & 49.103 & 48.184 & 46.964 & 49.091 & 47.282 & 43.971 & 42.848 \\ 
\hline
\simname{N0.1ZloSWR} & 50.253 & 49.595 & 50.145 & 46.134 & 49.525 & 50.054 & 44.074 & 42.060 \\ 
\simname{N0.5ZloSWR} & 49.929 & 49.374 & 49.787 & 46.251 & 49.323 & 49.697 & 44.039 & 42.180 \\ 
\simname{N5ZloSWR} & 49.502 & 49.099 & 49.284 & 46.224 & 49.073 & 49.157 & 44.000 & 42.396 \\
\simname{N30ZloSWR} & 49.084 & 48.653 & 48.882 & 46.295 & 48.623 & 48.701 & 43.819 & 42.516 \\ 
\simname{N100ZloSWR} & 48.858 & 48.416 & 48.664 & 45.923 & 48.392 & 48.397 & 43.743 & 42.380 \\ 
\hline
\end{tabular}
  \caption{Table of energies calculated from each simulation in the containing winds and photoionisation runs 2 Myr after the supernova. Labels as in table \protect\ref{supernova:processes:energytable}.}
\label{supernova:environment:energytable}
\end{table*}

\begin{table*}
\begin{tabular}{l c c c c c c c c c}
\textbf{Runs} & \textbf{t$_{f}$/Myr} & \textbf{R$_{tot}$} & \textbf{R$_{kin}$} & \textbf{R$_{th}$} & \textbf{R$_{turb}$} & \textbf{S$_{kin}$} & \textbf{B$_{th}$} & \textbf{S$_{mom,bulk}$} & \textbf{S$_{mom,turb}$} \\ 
\hline
\simname{N0.1ZsoSWR} & 2.176 & 50.015 & 49.676 & 49.750 & 47.927 & 49.665 & 49.682 & 44.016 & 43.037 \\ 
\simname{N0.5ZsoSWR} & 1.458 & 49.884 & 49.498 & 49.654 & 47.501 & 49.482 & 49.605 & 43.949 & 42.828 \\ 
\simname{N5ZsoSWR} & 0.861 & 50.047 & 49.501 & 49.901 & 47.353 & 49.476 & 49.882 & 43.991 & 42.708 \\ 
\simname{N30ZsoSWR} & 0.073 & 50.834 & 49.551 & 50.811 & 48.200 & 48.985 & 50.808 & 43.567 & 42.603 \\ 
\simname{N100ZsoSWR} & 0.084 & 50.673 & 49.331 & 50.653 & 47.887 & 49.050 & 50.650 & 43.648 & 42.667 \\ 
\hline
\simname{N0.1ZloSWR} & 3.694 & 50.424 & 50.003 & 50.217 & 48.474 & 49.961 & 49.878 & 44.389 & 43.337 \\ 
\simname{N0.5ZloSWR} & 2.461 & 50.096 & 49.740 & 49.844 & 48.019 & 49.728 & 49.675 & 44.316 & 43.166 \\ 
\simname{N5ZloSWR} & 1.401 & 50.022 & 49.543 & 49.846 & 47.346 & 49.530 & 49.776 & 44.289 & 42.972 \\ 
\simname{N30ZloSWR} & 0.622 & 50.204 & 49.349 & 50.139 & 47.575 & 49.302 & 50.108 & 44.159 & 43.003 \\ 
\simname{N100ZloSWR} & 0.228 & 50.488 & 49.287 & 50.460 & 47.936 & 49.108 & 50.445 & 44.019 & 42.903 \\ 
\hline
\end{tabular}
  \caption{Table of energies calculated from each simulation in the containing winds and photoionisation runs t$_{f}$ after the supernova. t$_{f}$ is defined by \protect\cite{Thornton1998} as 13 times the time at which the total luminosity via radiative cooling is at maximum. Labels as in table \protect\ref{supernova:processes:energytable}.}
\label{supernova:environment:energytable_tfinal}
\end{table*}

In this appendix we include sampled values for the energies and momenta in each run. In tables \ref{supernova:processes:energytable} and \ref{supernova:processes:energytable_tfinal} we give values for each of the runs at 0.1 and 30 \atcc that include a supernova only, a supernova and stellar winds, a supernova and photoionisation, and all three processes. In tables \ref{supernova:environment:energytable} and \ref{supernova:environment:energytable_tfinal} we give values for all runs that include all three processes, varying according to density and metallicity in the external medium. Values are given at at 2 Myr after the supernova, and at t$_f$. t$_f$ is  defined by \cite{Thornton1998} as 13 t$_0$, where t$_0$ is the time at which the luminosity from radiative cooling is at a maxmimum in the system as a whole. Unsurprisingly, the majority of the kinetic energy is in the shell. However, for late times the shell accounts for much of the thermal energy in the system, since the bubble has cooled rapidly from temperatures of $\sim 10^7$ K. By contrast, the high density of the shell allows it to retain a large amount of thermal energy even though its temperature is relatively low.

\end{document}